\documentclass[12pt]{iopart}

\usepackage{graphicx}
\usepackage{color}
\usepackage{epsfig}
\usepackage{float}
\usepackage{multirow}
\expandafter\let\csname equation*\endcsname\relax
\expandafter\let\csname endequation*\endcsname\relax
\usepackage{amsfonts}
\usepackage{amssymb}
\usepackage{amsmath}
\usepackage{tabularx,url,color}
\usepackage{hyperref}
\usepackage{ulem}
\usepackage{subfigure}
\usepackage{lineno}
\usepackage{upgreek}
\usepackage{comment}
\usepackage{xcolor}
\usepackage{tablefootnote}

\usepackage{orcidlink}  

\input{definitions}

\begin{document}
\leftline{Dated: \today}

\title{Calibration of the AdvancedVirgo+ Gravitational Wave Detector and Reconstruction of the Detector Strain $h(t)$ during the Observing Run O4}



\author{
F Acernese$^{2,3}$ %
A Agapito\orcidlink{0009-0005-9004-3163}$^{4}$ %
D Agarwal\orcidlink{0000-0002-8735-5554}$^{5}$ %
I-L Ahrend$^{6}$ %
L Aiello\orcidlink{0000-0003-2771-8816}$^{7,8}$ %
A Ain\orcidlink{0000-0003-4534-4619}$^{9}$ %
W Ali$^{10,11}$ %
A Allocca\orcidlink{0000-0002-5288-1351}$^{12,3}$ %
W Amar\orcidlink{0009-0003-5623-8819}$^{13}$ %
A Amato\orcidlink{0000-0001-9557-651X}$^{14,15}$ %
F Amicucci\orcidlink{0009-0005-2139-4197}$^{16,17}$ %
C Amra$^{18}$ %
M Andia\orcidlink{0000-0003-3675-9126}$^{19}$ %
T Andri\'c\orcidlink{0000-0002-9277-9773}$^{20,21}$ %
S Antier\orcidlink{0000-0002-7686-3334}$^{19}$ %
M Arca Sedda\orcidlink{0000-0002-3987-0519}$^{20,21}$ %
F Arciprete\orcidlink{0000-0003-3602-3717}$^{7,8}$ %
F Armato\orcidlink{0000-0002-8856-8877}$^{10,11}$ %
N Arnaud\orcidlink{0000-0001-6589-8673}$^{22}$ %
L Asprea$^{23}$ %
M Assiduo$^{24,25}$ %
S Assis de Souza Melo\orcidlink{0000-0002-1550-1671}$^{26}$ %
P Astone\orcidlink{0000-0003-4981-4120}$^{16}$ %
F Attadio\orcidlink{0009-0008-8916-1658}$^{17,16}$ %
F Aubin\orcidlink{0000-0003-1613-3142}$^{27}$ %
G Avallone\orcidlink{0000-0001-5482-0299}$^{2}$ %
N Avdeev\orcidlink{0009-0005-0413-633X}$^{23}$ %
S Babak\orcidlink{0000-0001-7469-4250}$^{6}$ %
S Bagnasco\orcidlink{0000-0001-6062-6505}$^{23}$ %
S Baimukhametova\orcidlink{0009-0006-0971-8619}$^{28,29}$ %
T Baka\orcidlink{0000-0002-5629-3813}$^{30,15}$ %
G Balbi$^{31}$ %
G Baldi\orcidlink{0000-0001-8963-3362}$^{32,33}$ %
N Baldicchi\orcidlink{0009-0009-8888-291X}$^{34,35}$ %
G Ballardin$^{26}$ %
M Ballelli\orcidlink{0000-0003-1512-5423}$^{20,21}$ %
B Banerjee\orcidlink{0000-0002-8008-2485}$^{20}$ %
M Baratti\orcidlink{0009-0003-5744-8025}$^{36,37}$ %
F Barone\orcidlink{0000-0002-8069-8490}$^{38,3}$ %
M Barsuglia\orcidlink{0000-0002-1180-4050}$^{6}$ %
D Barta\orcidlink{0000-0001-6841-550X}$^{39}$ %
A Basti\orcidlink{0000-0003-2895-9638}$^{37,36}$ %
M Bawaj\orcidlink{0000-0003-3611-3042}$^{34,35}$ %
M Bazzan$^{40,41}$ %
F Beirnaert\orcidlink{0000-0002-4003-7233}$^{42}$ %
M Bejger\orcidlink{0000-0002-4991-8213}$^{43}$ %
C Bellani\orcidlink{0000-0003-3267-1450}$^{44}$ %
D Beltran-Martinez\orcidlink{0000-0003-4580-3264}$^{45}$ %
E Benedetti\orcidlink{0009-0008-5230-0597}$^{16}$ %
I Bentara\orcidlink{0009-0000-5074-839X}$^{22}$ %
S Bera\orcidlink{0000-0003-0907-6098}$^{46}$ %
D Bersanetti\orcidlink{0000-0002-7377-415X}$^{10}$ %
T Bertheas\orcidlink{0009-0005-4118-4170}$^{47}$ %
A Bertolini$^{15,14}$ %
J Bezerra-Sobrinho\orcidlink{0000-0003-2183-4488}$^{48}$ %
V Biancalana\orcidlink{0000-0002-1642-5391}$^{49}$ %
F Bianchi$^{35}$ %
M Bilicki\orcidlink{0000-0002-3910-5809}$^{50}$ %
A Binetti\orcidlink{0000-0001-6449-5493}$^{44}$ %
S Biot$^{51}$ %
M Bitossi\orcidlink{0000-0002-9862-4668}$^{26,36}$ %
M-A Bizouard\orcidlink{0000-0002-4618-1674}$^{52}$ %
M Bloch$^{53}$ %
G Boileau\orcidlink{0000-0002-3576-6968}$^{52}$ %
M Boldrini\orcidlink{0000-0001-9861-821X}$^{26}$ %
R Bonnand\orcidlink{0000-0001-5013-5913}$^{13,54}$ %
N Borghi\orcidlink{0000-0002-2889-8997}$^{55,31}$ %
V Boschi\orcidlink{0000-0001-8665-2293}$^{36}$ %
Y Bothra\orcidlink{0000-0002-9380-6390}$^{15,56}$ %
A Boudon$^{22}$ %
A Bozzi$^{26}$ %
C Bradaschia$^{36}$ %
M Branchesi\orcidlink{0000-0003-1643-0526}$^{20,21}$ %
T Briant\orcidlink{0000-0002-6013-1729}$^{57}$ %
A Brillet\footnote{Deceased, March 2026.}$^{52}$ %
M L Brozzetti\orcidlink{0000-0002-5260-4979}$^{34,35}$ %
G Bruno$^{5}$ %
F Bucci\orcidlink{0000-0003-1726-3838}$^{25}$ %
A Buchicchio$^{17}$ %
A Buggiani$^{26}$ %
O Bulashenko\orcidlink{0000-0003-1720-4061}$^{58,59}$ %
T Bulik$^{60}$ %
H J Bulten$^{15}$ %
R Buscicchio\orcidlink{0000-0002-7387-6754}$^{61,62}$ %
N Busdon$^{40}$ %
% ...
{(The Virgo Collaboration)}
}
% ...
\address{$^{2}$Dipartimento di Fisica ``E.R. Caianiello'', Universit\`a di Salerno, I-84084 Fisciano, Salerno, Italy}
\address{$^{3}$INFN, Sezione di Napoli, I-80126 Napoli, Italy}
\address{$^{4}$Centre de Physique Th\'eorique, Aix-Marseille Universit\'e, Campus de Luminy, 163 Av. de Luminy, 13009 Marseille, France}
\address{$^{5}$Universit\'e catholique de Louvain, B-1348 Louvain-la-Neuve, Belgium}
\address{$^{6}$Universit\'e Paris Cit\'e, CNRS, Astroparticule et Cosmologie, F-75013 Paris, France}
\address{$^{7}$Universit\`a di Roma Tor Vergata, I-00133 Roma, Italy}
\address{$^{8}$INFN, Sezione di Roma Tor Vergata, I-00133 Roma, Italy}
\address{$^{9}$Universiteit Antwerpen, 2000 Antwerpen, Belgium}
\address{$^{10}$INFN, Sezione di Genova, I-16146 Genova, Italy}
\address{$^{11}$Dipartimento di Fisica, Universit\`a degli Studi di Genova, I-16146 Genova, Italy}
\address{$^{12}$Universit\`a di Napoli ``Federico II'', I-80126 Napoli, Italy}
\address{$^{13}$Univ. Savoie Mont Blanc, CNRS, Laboratoire d'Annecy de Physique des Particules - IN2P3, F-74000 Annecy, France}
\address{$^{14}$Maastricht University, 6200 MD Maastricht, Netherlands}
\address{$^{15}$Nikhef, 1098 XG Amsterdam, Netherlands}
\address{$^{16}$INFN, Sezione di Roma, I-00185 Roma, Italy}
\address{$^{17}$Universit\`a di Roma ``La Sapienza'', I-00185 Roma, Italy}
\address{$^{18}$Aix Marseille Univ, CNRS, Centrale Med, Institut Fresnel, F-13013 Marseille, France}
\address{$^{19}$Universit\'e Paris-Saclay, CNRS/IN2P3, IJCLab, 91405 Orsay, France}
\address{$^{20}$Gran Sasso Science Institute (GSSI), I-67100 L'Aquila, Italy}
\address{$^{21}$INFN, Laboratori Nazionali del Gran Sasso, I-67100 Assergi, Italy}
\address{$^{22}$Universit\'e Claude Bernard Lyon 1, CNRS, IP2I Lyon / IN2P3, UMR 5822, F-69622 Villeurbanne, France}
\address{$^{23}$INFN Sezione di Torino, I-10125 Torino, Italy}
\address{$^{24}$Universit\`a degli Studi di Urbino ``Carlo Bo'', I-61029 Urbino, Italy}
\address{$^{25}$INFN, Sezione di Firenze, I-50019 Sesto Fiorentino, Firenze, Italy}
\address{$^{26}$European Gravitational Observatory (EGO), I-56021 Cascina, Pisa, Italy}
\address{$^{27}$Universit\'e de Strasbourg, CNRS, IPHC UMR 7178, F-67000 Strasbourg, France}
\address{$^{28}$D\'epartement de Physique Nucl\'eaire et Corpusculaire, Universit\'e de Gen\`eve, 24 quai E. Ansermet, CH-1211 Geneva, Switzerland}
\address{$^{29}$Gravitational Wave Science Center, UniGe, -, Switzerland}
\address{$^{30}$Institute for Gravitational and Subatomic Physics (GRASP), Utrecht University, 3584 CC Utrecht, Netherlands}
\address{$^{31}$Istituto Nazionale Di Fisica Nucleare - Sezione di Bologna, viale Carlo Berti Pichat 6/2 - 40127 Bologna, Italy}
\address{$^{32}$Universit\`a di Trento, Dipartimento di Fisica, I-38123 Povo, Trento, Italy}
\address{$^{33}$INFN, Trento Institute for Fundamental Physics and Applications, I-38123 Povo, Trento, Italy}
\address{$^{34}$Universit\`a di Perugia, I-06123 Perugia, Italy}
\address{$^{35}$INFN, Sezione di Perugia, I-06123 Perugia, Italy}
\address{$^{36}$INFN, Sezione di Pisa, I-56127 Pisa, Italy}
\address{$^{37}$Universit\`a di Pisa, I-56127 Pisa, Italy}
\address{$^{38}$Dipartimento di Medicina, Chirurgia e Odontoiatria ``Scuola Medica Salernitana'', Universit\`a di Salerno, I-84081 Baronissi, Salerno, Italy}
\address{$^{39}$HUN-REN Wigner Research Centre for Physics, H-1121 Budapest, Hungary}
\address{$^{40}$Universit\`a di Padova, Dipartimento di Fisica e Astronomia, I-35131 Padova, Italy}
\address{$^{41}$INFN, Sezione di Padova, I-35131 Padova, Italy}
\address{$^{42}$Universiteit Gent, B-9000 Gent, Belgium}
\address{$^{43}$Nicolaus Copernicus Astronomical Center, Polish Academy of Sciences, 00-716, Warsaw, Poland}
\address{$^{44}$Katholieke Universiteit Leuven, Oude Markt 13, 3000 Leuven, Belgium}
\address{$^{45}$Centro de Investigaciones Energ\'eticas Medioambientales y Tecnol\'ogicas, Avda. Complutense 40, 28040, Madrid, Spain}
\address{$^{46}$Aix-Marseille Universit\'e, Universit\'e de Toulon, CNRS, CPT, Marseille, France}
\address{$^{47}$Laboratoire des 2 infinis - Toulouse, Universit\'e de Toulouse, CNRS/IN2P3, Toulouse, France, Toulouse, France}
\address{$^{48}$Federal University of Rio Grande do Norte, Campus Universit\'ario - Lagoa Nova, Natal - RN, 59078-970, Brazil}
\address{$^{49}$Universit\`a di Siena, Dipartimento di Scienze Fisiche, della Terra e dell'Ambiente, I-53100 Siena, Italy}
\address{$^{50}$Center for Theoretical Physics, Polish Academy of Sciences, 02-668, Warsaw, Poland}
\address{$^{51}$Universit\'e libre de Bruxelles, 1050 Bruxelles, Belgium}
\address{$^{52}$Universit\'e C\^ote d'Azur, Observatoire de la C\^ote d'Azur, CNRS, Artemis, F-06304 Nice, France}
\address{$^{53}$Subatech, CNRS/IN2P3 - IMT Atlantique - Nantes Universit\'e, 4 rue Alfred Kastler BP 20722 44307 Nantes C\'EDEX 03, France}
\address{$^{54}$Centre national de la recherche scientifique, 75016 Paris, France}
\address{$^{55}$DIFA- Alma Mater Studiorum Universit\`a di Bologna, Via Zamboni, 33 - 40126 Bologna, Italy}
\address{$^{56}$Department of Physics and Astronomy, Vrije Universiteit Amsterdam, 1081 HV Amsterdam, Netherlands}
\address{$^{57}$Laboratoire Kastler Brossel, Sorbonne Universit\'e, CNRS, ENS-Universit\'e PSL, Coll\`ege de France, F-75005 Paris, France}
\address{$^{58}$Institut de Ci\`encies del Cosmos (ICCUB), Universitat de Barcelona (UB), c. Mart\'i i Franqu\`es, 1, 08028 Barcelona, Spain}
\address{$^{59}$Departament de F\'isica Qu\`antica i Astrof\'isica (FQA), Universitat de Barcelona (UB), c. Mart\'i i Franqu\'es, 1, 08028 Barcelona, Spain}
\address{$^{60}$Astronomical Observatory, University of Warsaw, 00-478 Warsaw, Poland}
\address{$^{61}$Universit\`a degli Studi di Milano-Bicocca, I-20126 Milano, Italy}
\address{$^{62}$INFN, Sezione di Milano-Bicocca, I-20126 Milano, Italy}
% ...


\date{\today}

\begin{abstract}
From 10~April~2024 15:00~UTC to 18~November~2025 16:00~UTC, the AdvancedVirgo+ gravitational wave detector participated in the LIGO-Virgo-KAGRA O4 observing run, started on 24~May~2023~15:00~UTC. 
Around 173 transient gravitational wave (GW) sources, all corresponding to coalescences of binary compact objects involving black holes and neutron stars, were detected online during the two run periods O4b and O4c when Virgo was taking data in the detector network.
Despite its sensitivity being limited around 55~Mpc, the inclusion of Virgo into the network allowed to improve the accuracy of the source parameter estimation, in particular the sky localisation of the detected events.

This article describes the AdvancedVirgo+ detector calibration
and the re\-construction of the detector strain $h(t)$ during O4,
as well as the estimation of the associated frequency-dependent uncertainties.
The detector calibration is based on auxiliary actuators,
Newtonian Calibrators and Photon Calibrators, described in other publications. 

The $h(t)$~reconstruction, including linear noise subtraction, was processed online with a latency of about 10~s. The so-called AnalysisReady strain data were then produced offline. 
Most of the time, the strain time series was a copy of the online time series, but with updated frequency-dependent uncertainties, around 2-3\% in amplitude and below 30~mrad in phase in the 10-2000~Hz frequency band, with the exception of larger uncertainties around 50~Hz and 150~Hz.
The AnalysisReady strain data and associated uncertainties have been used for the offline LIGO-Virgo-KAGRA data analysis
and are also the data made publicly available through the Gravitational Wave Open Science Center (GWOSC).

\end{abstract}

\maketitle

\tableofcontents


\section{Introduction}
The fourth observing run (O4) of the LIGO-Virgo-KAGRA (LVK) collaboration started on 24 May 2023 and ended on 18 November 2025. The data from the different detectors of the network were used jointly to search for gravitational wave (GW) sources~\cite{Aasi:2013wya} in the frequency range 10~Hz to a few~kHz. Approximately 254 coalescences of binary compact objects (black holes or neutrons stars) were detected online with high significance during all O4~\cite{bib:GraceDB,bib:GWTC5,bib:GWOSCpaper}.

The run has been divided into three periods:  
O4a (24 May 2023 to 16 January 2024), 
O4b (10 April 2024 to 28 January 2025) 
and O4c (28 January 2025 to 18 November 2025), 
with a global detector commissioning break during O4c, from 1 April 2025 to 11 June 2025.
Data from the two Advanced LIGO interferometers~\cite{0264-9381-32-7-074001,PhysRevD.102.062003,bib:ligoreco}, 
located in USA, were available throughout the entire observing run.
Data from the \AdV\ interferometer~\cite{bib:2019_SQZ_O3_PhRvL.123w1108A,bib:2020_SQZ_source_galaxies8040079}, located in Europe (near Pisa, Italy), were available during the O4b and O4c periods, over which 173 online significant detections occurred.
In addition, data from the KAGRA interferometer~\cite{bib:KAGRA2021}, located in Japan, were available during various fractions of the run.
In this article, we focus on the data from the \AdV\ detector~\cite{TDR,TheVirgo:2014hva,TDR2019}.

\ \\
During~O4, the Advanced Virgo+ optical configuration was a double recycled interferometer with 3~kilometer long Fabry-Perot cavities in the arms~\cite{TDR,TheVirgo:2014hva}. One evolution, with respect to the O3 observing run (2019-2020), was the addition of the Signal Recycling (SR) mirror~\cite{bib:SignalRecycling} which introduced a second recycling cavity to increase the detector bandwidth.
However, because the Advanced Virgo+ recycling cavities are marginally stable,
this addition complicated the detector commissioning,
and the SR mirror was ultimately operated  with a slight misalignment during the O4 observing run in order to reduce the resonance of higher optical modes and optimize the sensitivity~\cite{bib:2026_DetectorPaper}.

\ \\
The GW strain modifies the length of the interferometer arms. Consequently, the differential arm-length variations induced by a passing GW produce power variations at the interferometer output port. 
However, in order to operate the detector, the relative positions of the various mirrors are precisely controlled~\cite{TDR} to maintain the different cavities in their nominal resonant conditions.
Because of the control bandwidth of about one hundred hertz, the interferometer response to gravitational waves is attenuated up to a few hundred hertz, and part of the GW information is no longer contained in the output power variations but instead in the mirror control signals.

\ \\
The main purpose of the Virgo calibration is to reconstruct the dimensionless detector strain $h(t)$, which describes the projection of the GW strain onto the Advanced Virgo+ interferometer, over the frequency range from 10~Hz to 10~kHz.
It is computed using the interferometer output signal and the longitudinal control signals.
The calibration requirements specified a maximum uncertainty of 7\% on the modulus of $h(t)$, 70~mrad on its phase around 100~Hz and 100~mrad on its phase around 2~kHz~\cite{bib:LIGO_requirements}.
Absolute timing is also a critical parameter to estimate the location of a GW source in the sky. Since the typical timing accuracy of GW searches is of the order of 0.1~ms~\cite{Localization:2018_LRR}, the absolute timing precision must be of the order of 0.01~ms or better~\cite{bib:2018CQGra..35t5004A}.
\ \\
In the long wavelength approximation~\cite{bib:Rakhmanov_2008_shortwavelengthapprox}, the differential arm length of the interferometer, $\Delta L(t) = L_x - L_y$, is related to the detector strain $h(t)$ by:
\begin{eqnarray}
h(t)=\frac{\Delta L(t)}{L_0}\quad\quad\text{where}\quad L_0\ = \ 3\ \text{km}\quad \text{is the Virgo arms length.}
\label{eq:1}
\end{eqnarray} 
For coherent GW searches with multiple detectors, the sign convention of $h(t)$ must be consistent across the detectors network. For Virgo, $L_x$ and $L_y$ denotes the length of the "North" and "West" arms, respectively.
\ \\
During O4, the principle of the $h(t)$ reconstruction remained the same as the one used during O2 and O3, and is described in~\cite{bib:2018CQGra..35t5004A, Accadia:2014fbz}: namely, the contribution of the control signals is added to the attenuated dark fringe output signal. This procedure requires to calibrate the responses of the mirror actuators, the response of the detection photodiodes readout electronics and to estimate the interferometer optical response (i.e. the power variation at the dark fringe output port induced by a given differential arm-length variation). 
The detector calibration relies on auxiliary mirror actuators that provide the length reference.
During O4, the primary length reference was, for the first time, a Newtonian calibrator (Ncal),
operating below a few tens of hertz. 
Photon calibrators (Pcal), taking Ncal as reference, has then been used to calibrate the Virgo detector up to a few kilohertz.
The description of these two types of reference actuators and their calibration during O4 are provided in~\cite{bib:2026_PCalO4,PhdGrimaud,bib:2020_PCalO3} and~\cite{bib:2024_NcalO4,bib:syx_thesis,bib:2020_NCalO3}.
\ \\

Two sets of Virgo strain data were produced for the O4 observing run.
First, the online $h(t)$ time series, distributed with a latency of about 10~s to the low-latency GW searches that generate public alerts for multi-messenger observatories. This online reconstruction was based on the calibration parameters estimated with the data taken before Virgo joined O4b. A preliminary estimate of the $h(t)$ uncertainties was distributed along with the online strain time series.

Second, the so-called Analysis Ready (AR) $h(t)$ time series was produced for the offline LIGO-Virgo-KAGRA (LVK) data analyses and was also released publicly through Gravitational Wave Open Science Center (GWOSC)~\cite{bib:GWOSC}.
In most cases, these strain time series were identical to the online versions, but with updated frequency-dependent uncertainty estimates.

\ \\
The purpose of this article is to provide an overview of the \AdV\ detector calibration and of the $h(t)$ strain reconstruction during the O4 observing run, together with their associated systematic uncertainties.
We focus primarily on the stability of the calibration models throughout O4
and on the estimation of the bias and uncertainties affecting the Analysis Ready $h(t)$ strain time series.
The results presented in this article correspond to the final detector calibration and uncertainty estimation, based on calibration data acquired during the whole run, and to the online or reprocessed reconstructed strain~$h(t)$.

\ \\
Section~\ref{sec:detector} briefly reviews the Advanced Virgo+ detector components that are relevant to calibration during O4. 
The calibration procedures are largely similar to those used during the O3 run~\cite{Virgocalib_2022}. Section~\ref{sec:calibo3o4} summarizes the main modifications done since then.
Sections~\ref{sec:calibTimingSensing} and~\ref{sec:calibActuators} present the results of the calibration of the photodiode readout chain and of the calibration of the mirrors actuation system. 

Section~\ref{sec:hrecAndNoiseSubtraction} describes how the $h(t)$ detector strain time series have been reconstructed using the relevant detector raw data and the associated calibration models. In addition, linear noise subtraction has been run to reduce noise from known sources in the $h(t)$ channel and thereby improve the Advanced Virgo+ sensitivity.

In Section \ref{sec:hrecerrors}, we describe the various validation checks done on the $h(t)$ channel and their use in estimating the systematic uncertainties affecting the amplitude, phase and timing of $h(t)$.

Finally, Section \ref{sec:ARframes} describes the production of Analysis Ready dataset containing the final calibrated $h(t)$ time series.


\section{The \AdV\ detector during the O4 run}
\label{sec:detector}

Most of the detector characteristics relevant to the calibration and reconstruction described
in~\cite{Virgocalib_2022} for O3 remain valid for the Advanced Virgo+ detector during O4.
They are briefly summarized in this section, with particular emphasis on the modifications introduced with respect to the O3 detector configuration.

\begin{figure}[!ht]
    \centering
	\includegraphics[trim={0 0cm 0 0cm},clip,scale=0.35]{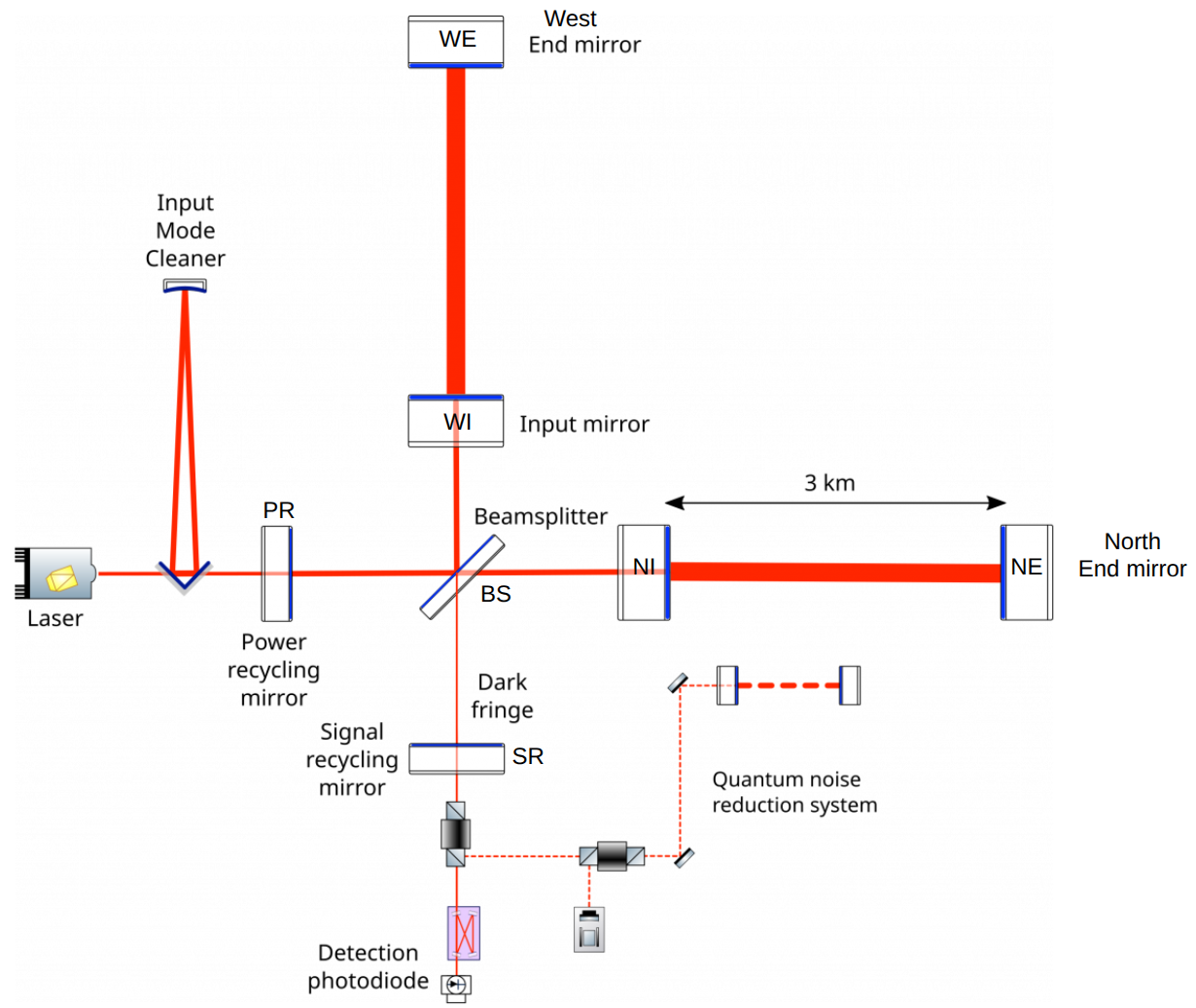} 
    \caption{Optical configuration of the Advanced Virgo+ interferometer for the O4 run.}
    \label{fig:itfconfig}
\end{figure}

The optical configuration of the Advanced Virgo+ detector during O4 is shown in Fig.~\ref{fig:itfconfig}: it is a power-recycled and signal-recycled interferometer with Fabry-Perot arm cavities as described in \cite{bib:2026_DetectorPaper}.
The finesse of the 3-km long Fabry-Perot cavities in the arms was approximately~450. 
The readout of the main interferometer output signal was based on homodyne (or DC) detection~\cite{TheVirgo:2014hva} 
and the photodiode signal used for the h(t) reconstruction was proportional to the interferometer differential arm length, commonly referred to as DARM. \\

In order to enable DC detection while maintaining destructive interference at the interferometer output port, the differential arm length was controlled with a small offset from the dark fringe operating point. A passing GW induces a small variation around this offset.\\


Data from the interferometer were sampled at 10~kHz for the control signals and 20~kHz for the output photodiode signals and were time-stamped using the Global Positioning System (GPS).\\

Each Virgo mirror is suspended to a complex seismic isolation system composed of a chain of seismic filters~\cite{bib:VirgoSuspension}.
The lower part of the suspension is a double-stage pendulum system with the so-called {\it marionette} as the first pendulum stage.
The mirrors are  suspended from the marionette using two pairs of thin fused-silica fibers~\cite{bib:AdvancedVirgoMonolithicPayload}. 
Since the suspension system is designed such that its resonances remain below 1~Hz, its mechanical response above 10~Hz has a simple $1/f^2$ behavior, which must be taken into account for the calibration.\\

The mirrors are shown in Fig.~\ref{fig:itfconfig} and are designated as BS for the Beam Splitter mirror, WI and NI for the West and North mirrors at the input of the arm cavities, WE and NE for the West and North mirrors at the end of the arm cavities, PR for the Power Recycling mirror and SR for the Signal Recycling mirror.

The positions of the marionette and mirror are controlled with electromagnetic actuators: permanent magnets are attached to the marionette and the rear surface of the mirrors while the corresponding coils are mounted on the reference cage suspended around each mirror~\cite{bib:2018_VirgoPayloads}. Electronic drivers are used to control the current flowing through every coil, thereby generating a magnetic field used to steer the suspended mirrors.

For the end mirrors, input mirrors and beam splitter mirror, the longitudinal controls are distributed between the marionette (up to a few tens of hertz) and the mirror (from approximately 10~Hz and up to a few hundred hertz).
Since the contributions of the various control signals are included in the detector strain reconstruction, the actuation responses of the NE, WE, NI, WI and BS marionette and mirrors must be measured up to about 100~Hz and 1000~Hz respectively. 

The principle of the longitudinal control loops is the same as described in~\cite{Accadia:2014fbz,bib:2020_ISC_O3_galaxies8040085}. 
The feed-forward technique implemented during~O3 to reduce the noise originating from the 50~Hz power mains was still used during O4. It consisted of measuring the transfer function between the DARM signal and one of the three phases extracted from the UPS \footnote{Uninterruptible Power Supply} monitoring and using it to modify the DARM control error signal. 
In contrast, the feed-forward technique setup in place during~O3 to mitigate the suspension cross-bar resonance at 48~Hz was no longer used during O4, since the motors that previously excited the cross-bars are now disconnected during data taking.


\section{Main evolutions of the Virgo calibration with respect to the O3 observing run}
\label{sec:calibo3o4}

The principle of the $h(t)$ reconstruction remained the same as during the O2 and O3 observing runs. Consequently,
the detector components requiring calibration are similar to those described in the O2 and O3 calibration articles~\cite{bib:2018CQGra..35t5004A,Virgocalib_2022}.
These are mainly the response of the output photodiode readout, 
the response of the electromagnetic actuators of the mirrors and marionette and the optical response of the interferometer.

The calibration procedures are also largely unchanged with respect to O3, although several improvements were introduced to reduce either the overall calibration time or some sources of uncertainty. The actuator calibration relies on calibration transfers from reference actuators:
by comparing the effect on the interferometer output power variations of mirror motions induced by a reference actuator and by the actuator to calibrate, one can derive the calibrated response of the latter.
Details of the calibration methods can be found in the O2 and O3 calibration articles~\cite{bib:2018CQGra..35t5004A,Virgocalib_2022}. 
In this article, we only briefly remind the principles and focus on the calibration results and
uncertainty estimation during the O4 observing run.

\subsection{Main relevant evolution of the Virgo detector}

The components requiring calibration during O4 were largely unchanged with respect to the O3 observing run. Nevertheless, new calibration measurements of all the actuators were performed during the months before the start of O4b.

The actuators of the arm-cavity mirrors were unchanged between the two runs O4b and O4c, but the WE mirror payload was replaced during the O4c commissioning break in spring 2025. Hence, for the last part of the O4c run, the WE mirror actuator response was re-calibrated and the associated calibration models were updated.

In order to frequency band where the interferometer is most sensitive, the SR mirror was added to the interferometer for O4, but its longitudinal control contribution was ultimately not included in the $h(t)$ reconstruction. Dedicated measurements were setup to calibrate the SR mirror actuators, but are not discussed in this article since they are not relevant to the $h(t)$ reconstruction.
Nevertheless, the addition of the SR mirror impacted the optical response of the interferometer. For instance,
the transfer function used to describe the interferometer optical response contains a simple pole,
whose frequency is approximately 55~Hz without SR and around 400~Hz with a tuned SR cavity.
In practice, the SR mirror was misaligned by $2\,\mu$rad to reduce the resonance of higher order optical modes in the marginally stable recycling cavities~\cite{bib:2026_DetectorPaper}, resulting in a cavity pole around 180~Hz during the O4 run.

The photodiodes at the output of the interferometer, together with the associated preamplifiers, were replaced between O3 and O4. The photodiodes installed in September 2023 were used during the first part of the run and were replaced again in September 2025, to test a design with lower electronic noise.
Dedicated measurements were made in September 2023 and in September 2025 to calibrate the photodiode sensing chain.

\subsection{Improved reference auxiliary actuators: Pcal and Ncal}

In order to calibrate the mirror actuators and to set the length reference for the $h(t)$ reconstruction, calibrated auxiliary reference actuators are used: 
the Newtonian calibrators~\cite{bib:2024_NcalO4,bib:syx_thesis,bib:2020_NCalO3} (Ncal), using a set of rotors to induce a variable gravity field around NE and WE mirrors, and the Photon calibrators~\cite{bib:2026_PCalO4, PhdGrimaud, PhdLagabbe} (Pcal), using auxiliary laser beams reflected on the NE and WE mirrors to apply a modulated radiation pressure on them.\\

The uncertainties of the Virgo Pcal calibration were reduced from 1.20\% during the O3 run~\cite{bib:2020_PCalO3} down to 0.48\% for the O4 run~\cite{bib:2026_PCalO4,PhdGrimaud,bib:calib_nist}. During O4, the Ncal system was operating continuously for the first time with two pairs of rotors located around the NE mirror. As a result, it had a calibration uncertainty of 0.17\% \cite{bib:2024_NcalO4} before the start of the run.
Therefore, it was decided to use the Ncals as the main length reference for O4. The resulting correction factors for the Pcals have then been respectively 1.00 for the WE Pcal and 1.09 for the NE Pcal. Once these factors were taken into account, the re-calibrated Pcals have then been used as reference to calibrate the mirror electromagnetic actuators over a large frequency band, from 10~Hz to 2~kHz. 
However, since the use of the Ncals, and in particular their stability, was not yet assessed when estimating the final calibration uncertainties, we have chosen to keep as calibration uncertainties the Pcal uncertainties and not the Ncal uncertainties.\\

\begin{figure}[!htb]
    \centering
    \includegraphics[trim={0 0cm 0 0cm},clip,scale=0.3]{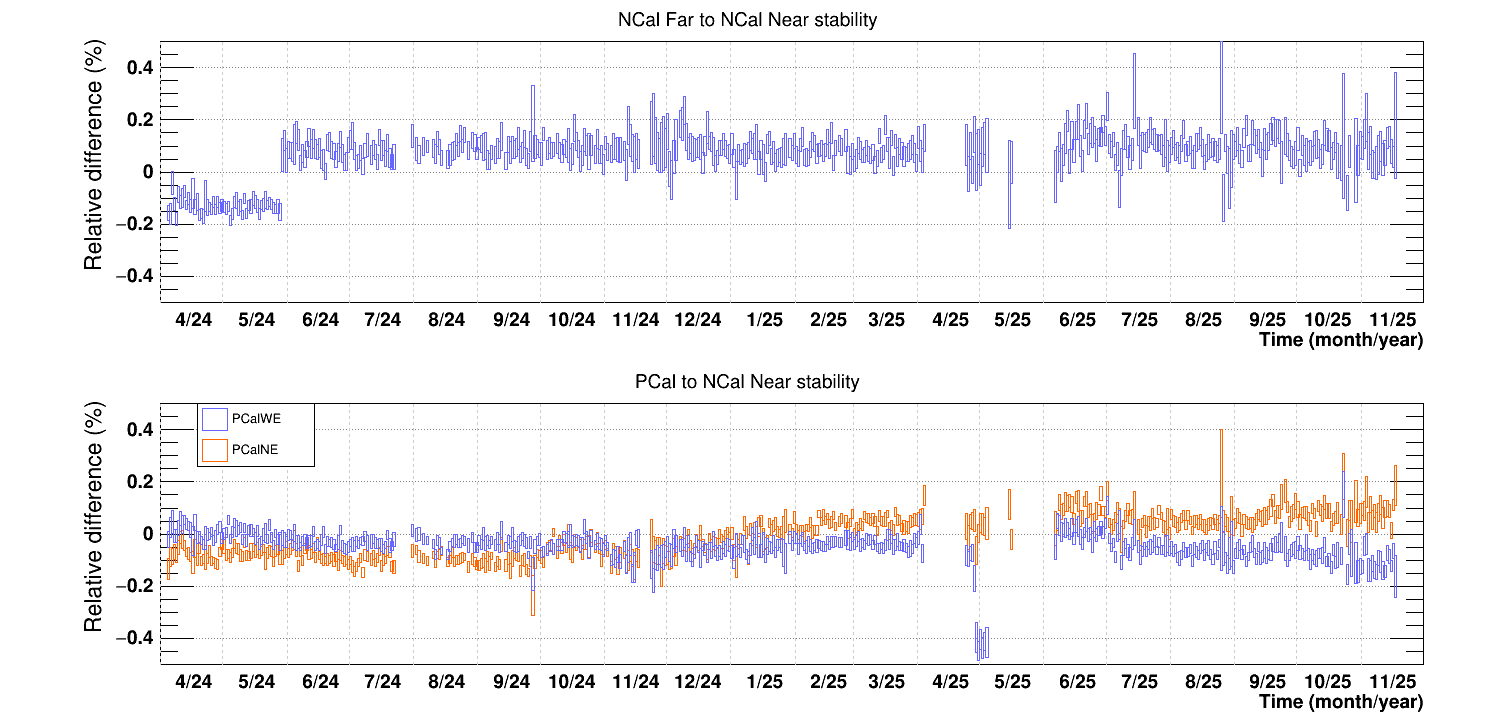}
    \caption{Relative stability of the Ncal and Pcal permanent calibration signals. The center of each bar in these plots is a mean value computed over one day. The vertical width of the bar is the statistical uncertainty for this one-day measurement. On 28 April 2025, to test the impact of the presence of point absorbers (which are small defects or impurities on the mirror that may locally introduce laser beam absorption), the WE mirror was displaced by 6~mm. This introduced a bias of 0.4\% on the WE Pcal until the WE mirror was changed in May 2025.}
    \label{fig:ncal_stability_o4}
\end{figure}

The relative stability of the permanent calibration signals (near 36~Hz) produced by the two pairs of Ncal and the two Pcals over the O4 run is presented in Fig.~\ref{fig:ncal_stability_o4}.
The top plot compares the two pairs of Ncal.
The jump observed a few weeks after the start of the run is due to a change of one pair of Ncals from having a rotor made of aluminum to a rotor made of PVC.
This change allowed a reduction of the parasitic coupling with the mirror control magnets due to Eddy current generated by the aluminum rotors spinning in the Earth magnetic field.
After this change, the Ncal systematic uncertainty was reduced to 0.12\%~\cite{bib:syx_thesis}, matching the observed stability and difference between the two pairs of Ncal.
The bottom plot shows the relative ratios between the calibration signal of the two Pcals and of the Ncals pair which was made of PVC from the start of the run and remained unchanged during O4.
Some small long-term drifts at the level of $\pm$~0.1\% are observed. Given the stability of the Ncal comparison, and the fact that the drifts are different for the two Pcal compared to one Ncal, such drifts are most probably coming from variations in the Pcal calibration~\cite{bib:2026_PCalO4}.

\subsection{Calibration data taking and continuous injection of sine-waves}
Most of the calibration data taking is based on applying motions to some of the mirrors of the interferometer.
The measurements can be split into two types: 
specific calibration measurements, out of run periods of the interferometer, where strong mirror motions are applied in sequence, for a few minutes each, either at specific frequencies, or as broad-band motions;
and mirror motion applied continuously at some particular frequencies, used for long-term monitoring
of the calibration and strain reconstruction.\\

Since the measurements done before and in the first weeks of O4b have shown a good stability of the calibration parameters for the mirror actuators, the periodicity of the measurements for mirror actuator calibration has been reduced to a bi-weekly cadence. Some dedicated measurements used to estimate the bias of the reconstructed detector strain time series $h(t)$ were still run on a weekly basis during O4. Overall, the fraction of calibration time over O4 has been 1.7\%.\\

As for the past Virgo runs, two sets of 3 permanent sine-wave injections moving
the NE, WE and BS mirrors were applied continuously
to monitor the slow variations of relevant parameters for the reconstruction
of the detector strain $h(t)$ (see section~\ref{sec:hrecOR}).

In addition, as already introduced during the O3 run, a set of permanent sine-wave injections moving the two end mirrors at fixed frequencies and amplitudes was applied over the whole O4 run to continuously monitor the calibration and to estimate the reconstructed strain bias and uncertainties (see section~\ref{sec:hrecerrors}).

For the O4 run, the full frequency range was covered using only the NE electromagnetic actuator,
to prevent additional uncertainties from using multiple actuators.
Some additional sine-waves were injected through the WE electromagnetic actuator and the NE and WE Pcal actuators as cross-checks.
In addition, continuous sine-waves were also injected by the Ncal around 36~Hz to monitor the cross-calibration of all the actuators around this frequency.



\section{Calibration of Advanced Virgo+ timing and sensing}
\label{sec:calibTimingSensing}

The main output signal of the interferometer is the power at the dark port,
provided as a time series, $\mathcal{P}_{DC}$, sampled at 20~kHz. 
It is sensed using two photodiodes whose signals are summed in the $h(t)$ reconstruction processing.
This processing has to take into account what we call the output photodiode sensing, which is a description of how the output DC power of the dark fringe beam is converted by the detection photodiodes into the $\mathcal{P}_{DC}$ time series.
It covers the analog preamplifier (low-pass filter around 40~kHz until September 2025, and around 400~kHz afterwards),
the analog anti-alias filter at the input of the ADC channel (low-pass filter around 80~kHz)
and the digital processing and filtering, ending with a 8th order low-pass Butterworth filter at 10~kHz.

An important parameter of this model is its overall delay, since the recorded time series is one of the main signals used to reconstruct the detector strain $h(t)$ time series that needs to be precisely synchronized with the GPS time.
Below a few kilohertz, the calibration sensing has been modeled during O4 by a 8th order low-pass Butterworth filter at 10~kHz, with a pure delay as an effective simple model (accurate within 0.01\% in modulus and 0.4\,\mus\ up to 10~kHz) for the digital data transmission and the filters at higher frequencies.

It also corrects for the 16\,\mus\ delay of the Virgo time series timestamp compared to GPS time, due to the centralized Virgo timing distribution.
The delay has been measured~\cite{bib:2026_Vnote_sensingO4} by flashing a 1~PPS signal~(Pulse Per Second signal, which is a synchronization signal delivered by the GPS receiver) using a LED, connected to a GPS receiver, temporarily put in front of the photodiodes.
It was measured after each photodiode installation, first in September 2023 then in September 2025.
The measured delays are in agreement within 1\,\mus\ with the expected delays from the full modelling of the sensing chain, which is $67$\,\mus\ until September 2025 and $63$\,\mus\ afterwards.
These measurements indicate also that the total uncertainty on the photodiode readout chain is of the order of 1\,\mus.

The data from September 2025 were not analysed quickly enough to update the calibration model used in the online $h(t)$ reconstruction process for the last part of O4c, which has introduced a bias of 4\,\mus\ in the raw reconstructed time series.
However, this bias was corrected within the online Hrec process, as described in Section \ref{sec:hrecerrors_weeklyO4}. 
Hence this error in the photodiode sensing modeling for the last two months of O4c had no impact
on the $h(t)$ time series that has been distributed to the low-latency data analyses or later in the AR dataset~\cite{bib:2026_Vnote_sensingO4}.

\section{Calibration of the electromagnetic actuators using the photon calibrators}
\label{sec:calibActuators}
The calibration of Advanced Virgo+ mirror and marionette electromagnetic actuators consists in measuring their transfer function (modulus in m/V) that converts a known digital signal (in volts) into a mirror displacement (in meters).
The measurement of this transfer function $A_{cur}$ is done using a reference for which the transfer function $A_{ref}$ is known. During O4, we used the Pcal as reference actuator to calibrate the interferometer actuators in the frequency range from 10~Hz to 2~kHz~\cite{bib:2026_PCalO4}.\\

When injecting a signal with the Pcal, the output signal of the interferometer $\mathcal{P}_{DC}^{ref}$ can be written as a combination of the injected signal $I_{ref}$, the Pcal actuator model $A_{ref}$ and the interferometer response $R$: $\mathcal{P}_{DC}^{ref} =I_{ref} \times A_{ref} \times R$.
When injecting a signal with the actuator to be calibrated, $A_{cur}$, the same combination applies: $\mathcal{P}_{DC}^{cur} = I_{cur} \times A_{cur} \times R$.
The actuator response of the actuator to be calibrated is estimated from two measured transfer functions and the knowledge of the reference actuator response:
\begin{equation}
    A_{cur} = \left(\frac{\mathcal{P}_{DC}^{cur}}{I_{cur}}\right) . \left(\frac{\mathcal{P}_{DC}^{ref}}{I_{ref}}\right)^{-1} \times A_{ref}
\label{eq:calibproc}
\end{equation}

This is the general method for actuator calibration that we apply to all mirror and marionette actuators~\cite{Virgocalib_2022}. 
It assumes that the response $R$ of the interferometer is the same during the two datasets (``ref" and ``cur"), taken a few minutes apart. Performing such measurements repeatedly allows to validate this hypothesis and to include possible small variations in systematic uncertainties.\\

\subsection{From the Pcal reference to the different mirror actuators}
The measurements to transfer the calibration from the Pcal actuators to the mirror electromagnetic actuators are very similar to what was described in more details in the O3 calibration article~\cite{Virgocalib_2022}. Only a brief summary is given here.

The Pcal calibration is directly transferred to the mirror and marionette electromagnetic actuators of the NE, WE, NI and WI mirrors $ ^{\footnote{While during O3, the NI and WI actuators were calibrated with an extra-step, using the NE and WE electromagnetic actuators as references.}}$ with the interferometer locked in its standard operating mode. 
In practice, as described in~\cite{Virgocalib_2022} $ ^{\footnote{Section~5.4.1 of~\cite{Virgocalib_2022}.}}$, the response $R$ of the interferometer
is slightly different for the motions of the inputs mirrors than for motions of the end mirrors. This is taken into account in the calibration transfer.

Then to calibrate the actuators of the BS mirror, the interferometer is put in a different configuration where
only the cavity formed by the WI, BS and PR mirrors is locked, the other mirrors being misaligned. From this configuration, the BS mirror electromagnetic actuator is calibrated
using the WI mirror actuator as reference. Finally, the BS marionette actuator is calibrated
using the BS mirror actuator as reference, with the interferometer locked in its standard operating mode.

\subsection{Calibration models estimated before the start of O4b}
In order to provide accurate actuators models for the $h(t)$ reconstruction during the O4 run, the different measurements associated to these calibration transfers have been done on a weekly basis from August~2023 to April~2024, before Virgo joined O4b. These measurements were used to estimate the calibration models
that have then been used during O4b and O4c for the $h(t)$ reconstruction,
except for the WE mirror actuator that was changed in June 2025.
For the last part of O4c, the WE actuator model was updated following measurements
done in June and beginning of July 2025.
The models used for the NE and WE mirror actuators in the $h(t)$ reconstruction 
are given in Table~\ref{tab:mirNE_WE_models}.
The parameters of the actuators models are a gain, a delay and a set of ad-hoc poles and zeros whose initial values are based on the knowledge of the electronics of the actuators and the mechanics of the mirror suspensions. More details about those models are provided in the appendix~\ref{sec:actmodels}.

\begin{table}[tbh!]
\begin{center}
\begin{tabular}{|c|c|c|c|}
\hline
Model               & O4 online NE & \multicolumn{2}{|c|}{O4 online WE} \\ 
                    &              & until May 2025    & from June 2025 \\
\hline
\bf{From fit}  &   & & \\
Gain ($\upmu$m$/$V) & $0.434$ & $0.409$  & $0.426$\\  
Pole $f_{p}$ (Hz)   & $105.5$ & $193.5$  & $137.2$\\ 
Zero $f_{z}$ (Hz)   & $110.1$ & $197.7$  & $139.4$\\ 
Zero $f_{z}$ (Hz)   & $6648.0 $ & $5269.0 $  & $8805.6$\\
Delay ($\upmu$s)   & $-157.6$  & $-155.3$   & $-163.4$\\
\hline
\bf{Pendulum} &   \multicolumn{3}{|c|}{} \\
$f_{0}$ (Hz) & \multicolumn{3}{|c|}{$0.6$} \\
$Q_{0}$      & \multicolumn{3}{|c|}{$1000$} \\
\hline
\end{tabular}
\caption{
\label{tab:mirNE_WE_models} 
NE and WE mirror electromagnetic actuators models used for the $h(t)$ reconstruction during~O4. The models ``O4 online" were derived using the measurements done between August 2023 and April 2024.  For WE, new measurements have been done and the actuator model has been updated in June 2025 because the WE mirror had been changed between May and June 2025.
The validity range of the models is from 10~Hz to 1500~Hz.
As shown on Fig.~\ref{fig:hrec_breakdown}, above 1000~Hz, the contribution of actuators models is negligible with respect to the photodiode signal and thus does not contribute to $h(t)$.
}
\end{center}
\end{table}

\subsection{Stability of the actuator calibration during O4}

\begin{figure}[tbh!]
    \centering
	\includegraphics[trim={0 0cm 0 0cm},clip,scale=0.4]{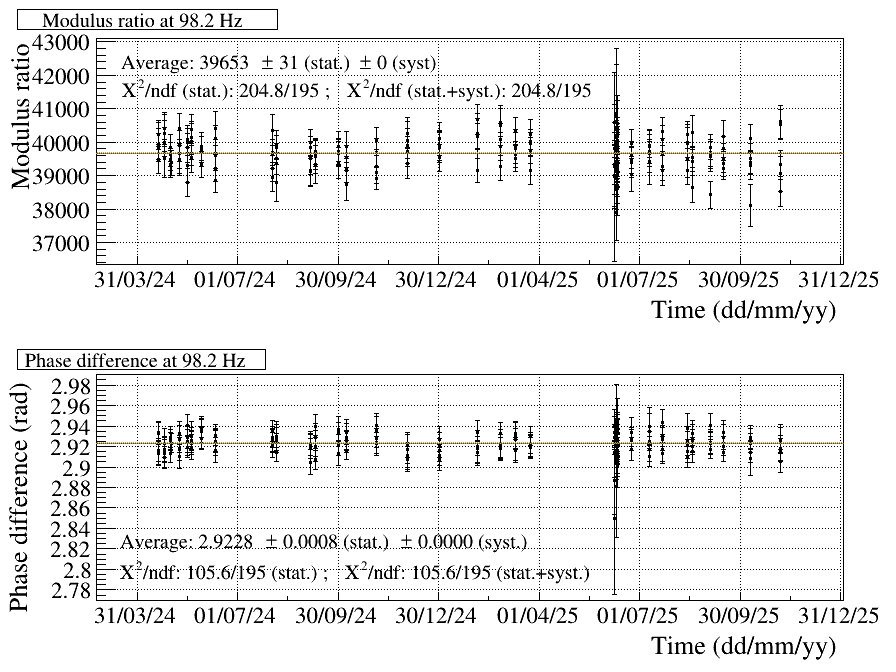} 
    \caption{Modulus and phase of the transfer function between the NE mirror actuator and the NE Pcal measured at the frequency of 98.5~Hz: regular measurements during O4b and O4c, including some post-O4c data.
    The average values are given with their statistical uncertainties and ${\chi}^2/\text{ndf}$.
    At this particular frequency, no additional systematic uncertainty is needed.
    }
    \label{fig:NEoverPCal98}
\end{figure}

\begin{figure}[tbh!]
    \centering
	\includegraphics[trim={0 0cm 0 0cm},clip,scale=0.3]{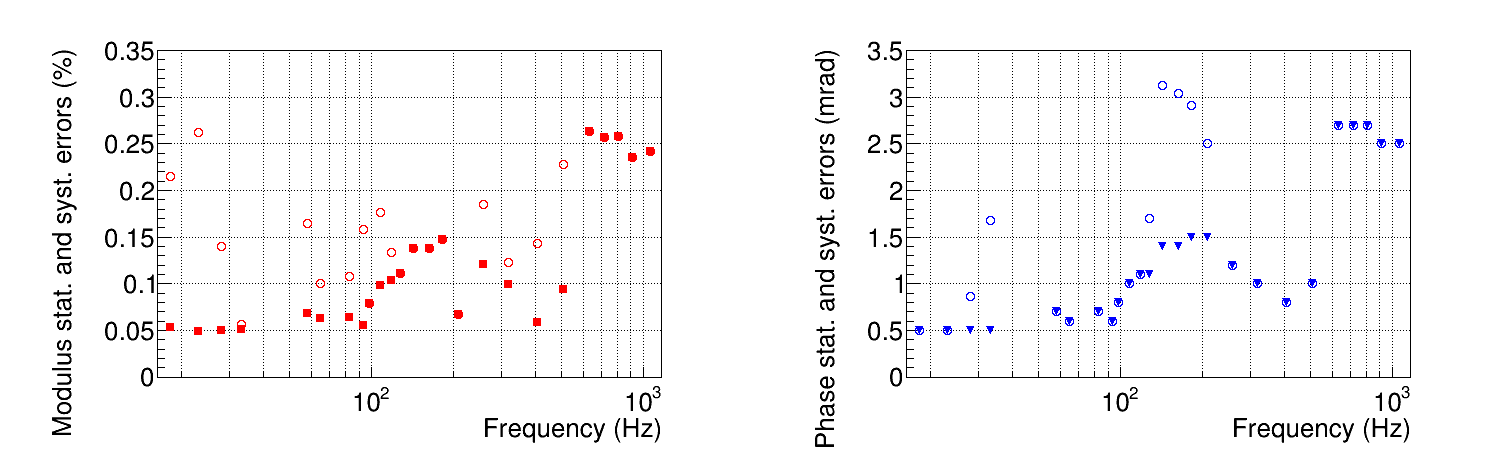} 
    \caption{Statistical and systematic errors estimated over O4b and O4c for the transfer function from the Pcal to the NE mirror actuator. The errors on the amplitude are expressed in [\%] and the errors on the phase are expressed in [mrad]. 
    Left: Statistical errors on the amplitude (red filled squares) of the calibration transfer
    as a function of frequency and statistical plus systematic errors (red empty circles).
    Right: Statistical errors on the phase (blue filled triangles) of the calibration transfer as a function of frequency and statistical plus systematic errors (blue empty circles).
    }
    \label{fig:NEoverPCal_uncertainties}
\end{figure}

During the run, the measurements were done on a bi-weekly basis in order to monitor the stability of the actuator calibration, while keeping low the associated downtime. 
Figure~\ref{fig:NEoverPCal98} shows the evolution, over the O4b+O4c run, of the modulus and phase of the transfer function between the NE mirror actuator and the NE photon calibrator at 98.2~Hz.
The line injected at this frequency is a monitoring one which was injected in all the data sets, whatever the calibration measurement done. The data shown on Fig.~\ref{fig:NEoverPCal98} is consistent with statistical variations, providing uncertainties below 0.1\% on the average modulus and 1~mrad on the average phase.
At frequencies where time variations are not compatible with statistical variations, an additional systematic uncertainty is estimated by an iterative method, increasing the error bars of every data points until the $\chi^2$ probability of the hypothesis saying that the data, with their increased error bars, are compatible with a constant value is at least 5\%. Figure~\ref{fig:NEoverPCal_uncertainties} summarizes the statistical and total uncertainties obtained at the frequencies that were measured. All total uncertainties are lower than 0.3\% in modulus and 3~mrad in phase.
As a consequence, the mirror actuator responses have been estimated using the average of all the calibration periods, using as error bars the total uncertainties described above. 

\newpage

\begin{figure}[tbh!]
    \centering
    \includegraphics[trim={0 0cm 0 0cm},clip,scale=0.35]{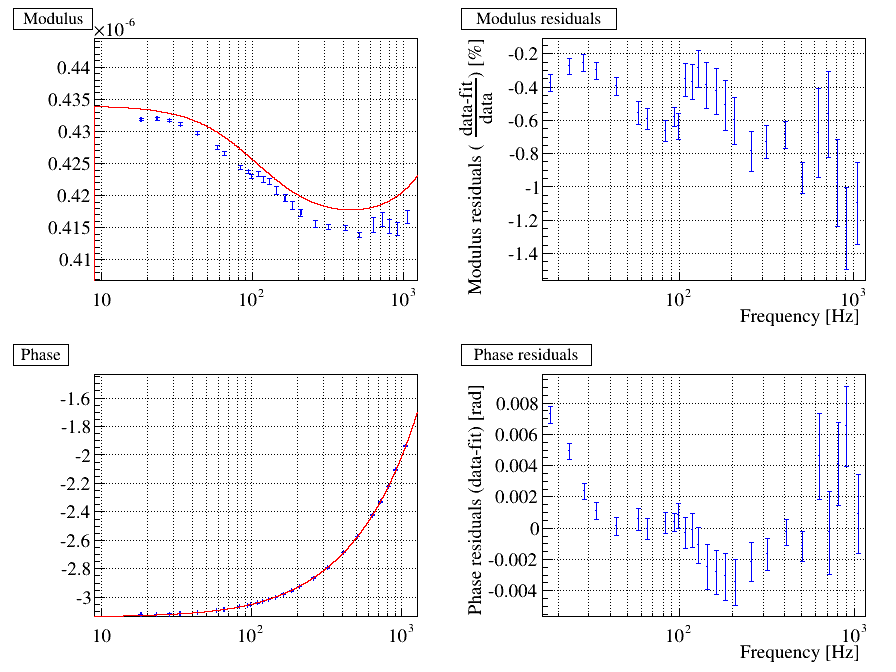}
    \caption{Measurements of the modulus and phase of the NE mirror actuator response, averaged over the time period 10 April 2024 (start of O4b) to 18 November 2025 (end of O4c). The superposed red line is the NE actuator response model used for the $h(t)$ reconstruction processing during that same period, that was estimated with pre-O4b data (see Table~\ref{tab:mirNE_WE_models}). To visualize only the electronic part of the actuator's model, the data have been divided by a second order low-pass filter at 0.6~Hz with a quality factor of 1000, which is the mechanical part of the actuator's model (see Table \ref{tab:mirNE_WE_models}).
    The plots on the right side show the residuals between the measurement and the model.}
    \label{fig:NEcalib_fitwithHrecmodel}
\end{figure}

\newpage

\begin{figure}[tbh!]
    \centering
	\includegraphics[trim={0 0cm 0 0cm},clip,scale=0.35]{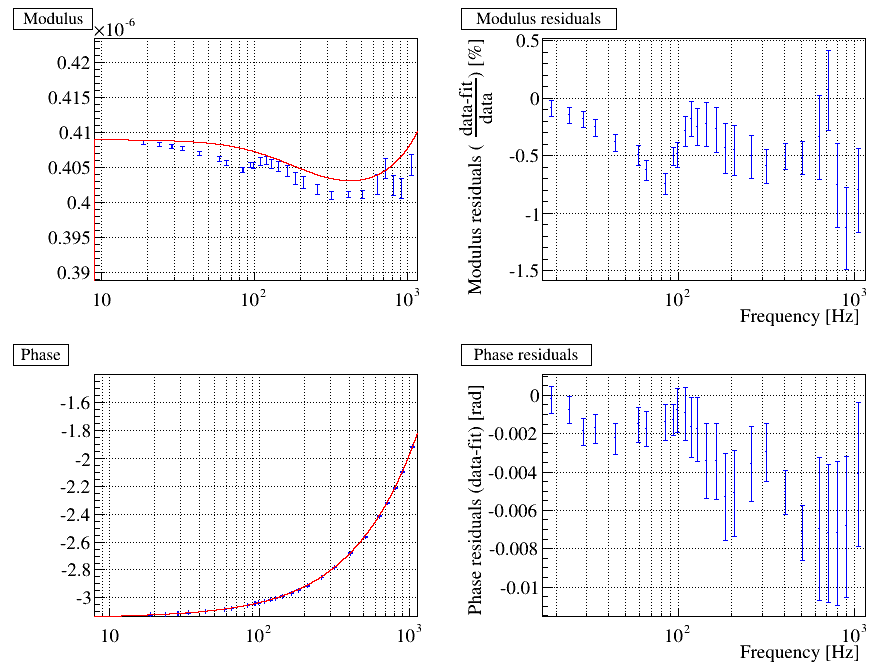}
    \caption{
    Measurement of the modulus and phase of the WE mirror actuator response averaged over the time period 10 April 2024 (start of O4b) to 10 April 2025 (before the WE mirror was replaced). The superposed red line is the WE actuator response model used for the $h(t)$ reconstruction processing during this same period, that was estimated with pre-O4b data (see Table~\ref{tab:mirNE_WE_models}). For visualization, the data have been normalized by a second order low-pass filter at 0.6~Hz with a quality factor of 1000, which is the mechanical part of the actuator's model (see Table \ref{tab:mirNE_WE_models}).
    The plots on the right side show the residuals between the measurement and the model.
    }
    \label{fig:WEcalib_fitwithHrecmodel}
\end{figure}

Figures~\ref{fig:NEcalib_fitwithHrecmodel} and~\ref{fig:WEcalib_fitwithHrecmodel} show, on the left side, the responses measured during O4b and O4c as blue points.
The red lines represent the calibration models estimated on the data from August 2023 to April 2024, before the start of O4b, and which are given in the Table~\ref{tab:mirNE_WE_models}.

On the right side, the residuals between the data taken during O4b and O4c and the model estimated with the data before the start of the run are shown: some frequency-dependent deviations are visible. 
From these deviations, a systematic uncertainty of 0.5\% has been estimated on the modulus of the actuator models. On the timing, the agreement is at the 1\,\mus\ level. This uncertainty is included in Table \ref{tab:MirUncertainties} in the line "O4 vs pre-O4 models".

\subsection{Detector calibration uncertainties estimation}
\label{sec:calibuncertainties}

Each step of the calibration procedure described in the previous Section and in~\cite{Virgocalib_2022} contributes 
to the total uncertainty of the models describing the actuator responses
that have been used in the $h(t)$ reconstruction processing during~O4.
Tables~\ref{tab:MirUncertainties} and~\ref{tab:MarUncertainties} provide the breakdown of the various contributions and the total uncertainty, respectively on the mirror and marionette actuator models.

\newpage

\begin{table}[tbh!]
\begin{center}
\small{
\begin{tabular}{|c|c|c|c|}
\cline{2-4}
\multicolumn{1}{c|}{}               & NE mirror       & WE mirror        &  BS mirror \\
\hline
  Pcal calibration                  & \multicolumn{3}{c|}{0.5\% ; 3 \mus}  \\
\hline
  Pcal to end transfer              & \multicolumn{2}{c|}{0.3\% ; 3 mrad} & --  \\ 
\hline
  WI to BS transfer                 & \multicolumn{2}{c|}{--}             & $<$ 1\% ; 10 mrad  \\
 \hline
 O4 vs pre-O4 models               & \multicolumn{3}{c|}{0.5\% ; 1 \mus}   \\
\hline
\hline
Total uncertainty                   & 0.77\%          & 0.77\%           & 1.3\%      \\
(quadratic sum)                     & 3.0~mrad          & 3.0~ mrad           & 10~mrad    \\
                                    & 3.2~\mus         & 3.2~\mus         & 3.2~\mus      \\
\hline                                                                                                            
Validity range                       & 10-1500 Hz     & 10-1500 Hz      & 10-1000 Hz   \\
\hline
\end{tabular}
    \caption{Summary of the uncertainties on the mirror actuator models.
  For every mirror actuator, the uncertainties on the modulus and phase (or timing when more relevant) of the model are given for the different contributions along the calibration procedure. Their quadratic sum gives the total uncertainties.
  Their validity range is also provided. The validity range does not span the full range of strain frequencies but is enough, given the unity gain frequency (about one hundred hertz) of the different loops acting on the mirror positions.
    }
    \label{tab:MirUncertainties}
}
\end{center}
\end{table}

\begin{table}[tbh!]
\begin{center}
\small{
\begin{tabular}{|c|c|c|c|}
\cline{2-4}
\multicolumn{1}{c|}{}                & NI mario.      & WI mario.    & BS mario.     \\  
\hline
  Pcal calibration                   & \multicolumn{3}{c|}{0.5\% ; 3 \mus}           \\
\hline
  Pcal to marionette           & 0.6\% ; 5 mrad & 3\% ; 5 mrad & 0.3\% ; 2 mrad \\  
  Model fit residuals                & 5\% ; 20 mrad  & 4\% ; 5 mrad & 2.5\% ; 5 mrad \\  
\hline           
\hline
\multicolumn{1}{|c|}{Total uncertainty} & 5.1\%       & 5.1\%        & 2.6\%           \\ 
\multicolumn{1}{|c|}{(quadratic sum)}  & 21 mrad     & 7.0 mrad       & 5.4 mrad         \\ 
\multicolumn{1}{|c|}{}                  & 3~\mus      & 3~\mus       & 3~\mus          \\ 
\hline                                      
\multicolumn{1}{|c|}{Validity range}    & 10-80 Hz    & 10-80 Hz     & 10-100 Hz       \\ 
\hline                                                                               
\end{tabular}
\caption{Summary of the uncertainties on the marionette actuator models, along with their validity range.
The validity range does not span the full range of strain frequencies but is enough, given the unity gain frequency (a few tens of hertz) of the different loops acting on the marionette positions.
}
\label{tab:MarUncertainties}
}
\end{center}
\end{table}

\section{Reconstruction of the detector strain h(t) and noise subtraction}
\label{sec:hrecAndNoiseSubtraction}
The Advanced Virgo+ detector strain time series $h(t)$ is reconstructed using both the raw interferometer data (photodiodes and mirror controls signals)
and the calibration models describing the photodiode sensing chain and the mirror actuators. 
This Section provides an overview of the strain reconstruction process,
while the associated uncertainty estimation for the O4 observing run is described in section~\ref{sec:hrecerrors}.
A brief description of the $h(t)$ reconstruction itself and of the unbiasing procedure implemented for O4 is given in section~\ref{sec:hrec}.
The adjustment of the optical response is described in section~\ref{sec:hrecOR}. 
The noise subtraction techniques used during O4 are briefly presented in section~\ref{sec:NoiseSubtraction}. 
Finally, some performance metrics of the $h(t)$ reconstruction during O4 are presented in section~\ref{sec:Hrecperf}.

\subsection{Principle of the h(t) reconstruction}
\label{sec:hrec}

As for the previous observing runs O2 and O3, the $h(t)$ reconstruction is computed in the frequency domain, using Fast Fourier Transforms (FFTs) computed over 8~s time windows with 4~s overlap~\cite{bib:TDSHrec, bib:Hrec_O4, bib:hrec_Amaldi}.

\begin{figure}[tbh!]
    \centering
	\includegraphics[trim={0 0cm 0 0cm},clip,scale=0.55]{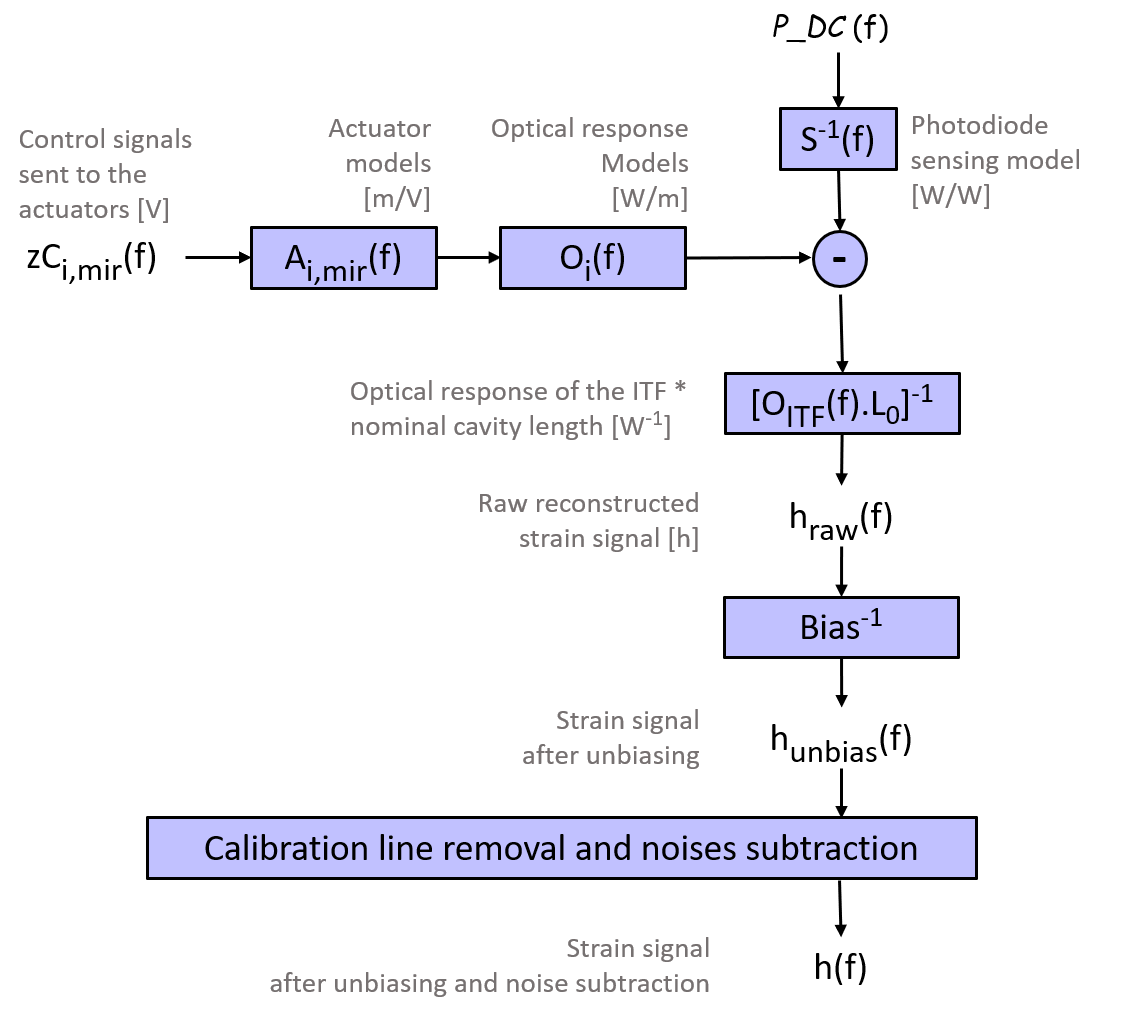} 
    \caption{
    Principle of the Virgo  detector strain reconstruction during the O4 observing run.
    The first output is the raw channel $h_{raw}$.
    A correction for the estimated bias is then applied to obtain the intermediate channel $h_{unbias}$
    Calibration lines are subsequently removed and linear noise subtraction is applied to produce the final channel $h(t)$.
    $\mathcal{P}_{DC}$ denotes the dark fringe output power of the interferometer.
    $zC_{i,m}$ is the control signal applied to the mirror or marionette actuator~$i$, whose response model is $A_{i,m}$.
    $O_i$ is the model of the interferometer optical response to a motion of mirror~$i$ and $O_{ITF}$ is the optical response to a passing gravitational wave.
    $L_0 = 3\,\text{km}$ is the length of the Virgo arm cavities.
    From the quantities $h_{raw}(f)$, $h_{unbias}(f)$ and $h(f)$, an inverse Fourier Transform produces the time series $h_{raw}(t)$, $h_{unbias}(t)$ and $h(t)$, all sampled at 20~kHz. 
    In parallel, it produces the main $h(t)$ time series sampled at 16384~Hz, which is the one used by the online and offline data analysis pipelines.
    }
    \label{fig:hrecprinciple}
\end{figure}

In the first stage of the reconstruction, the contributions of the control signals are removed from the dark-fringe signal by taking into account the actuators models and the interferometer optical response for each mirror or marionette, as illustrated in Fig.~\ref{fig:hrecprinciple}.
The input channels are the dark fringe photodiode signal $\mathcal{P}_{DC}$ and the control signals $zC_i$ sent to the $i^{th}$ mirror and marionette actuators.\\

\begin{figure}[tbh!]
    \centering
	\includegraphics[trim={0 0cm 0 0cm},clip,scale=0.55]{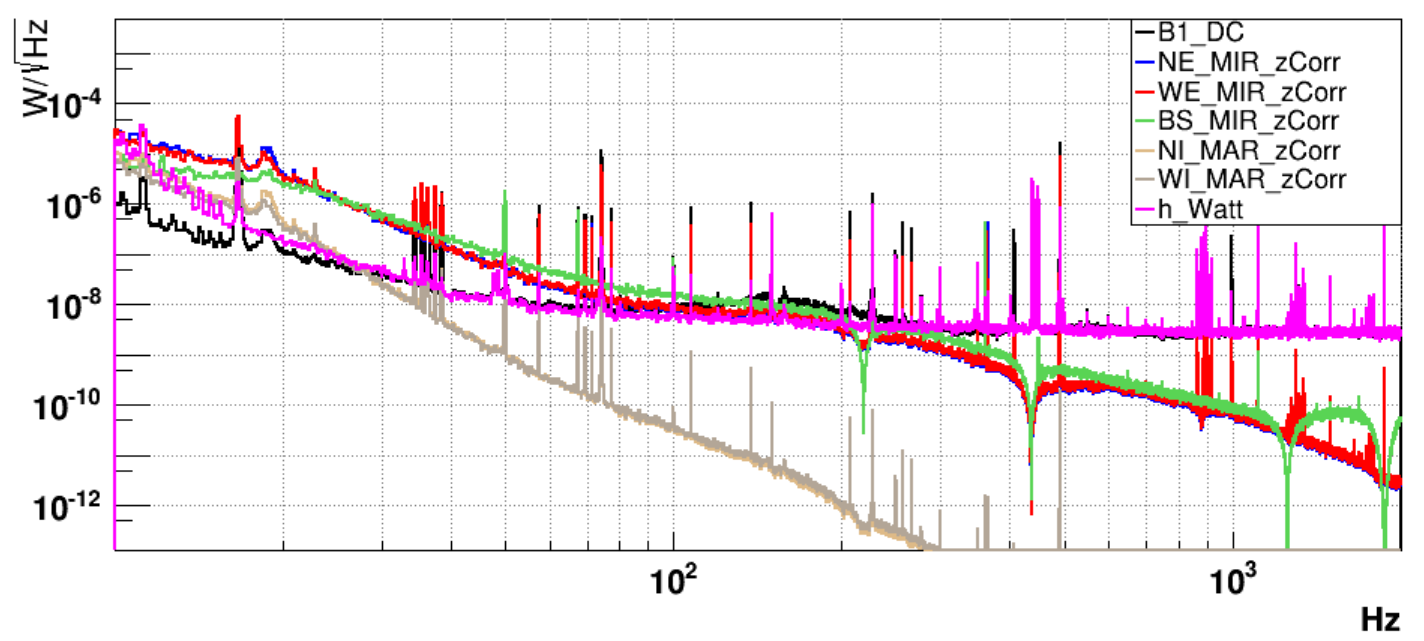} 
    \caption{Contributions of the photodiode output signal and of the mirror longitudinal control signals to the raw reconstructed detector strain channel~$h_{raw}$. 
    The pink curve shows the amplitude spectral density of $h_{raw}$ in equivalent watt units (that means after multiplication by the interferometer mean optical response). 
    The black curve shows the amplitude spectral density of the dark fringe signal~$\mathcal{P}_{DC}$.
    The other curves represent the contributions of the various control signals $zC_i$ in the same units. 
    Plot has been done with Virgo data at GPS=1441932000 (15~September 2025).
    }
    \label{fig:hrec_breakdown}
\end{figure}

In a first step, the calibrated dark-fringe power is obtained by applying the inverse transfer function of the photodiode readout sensing chain $S(f)$ to the signal $\mathcal{P}_{DC}$. Each control signal $zC_i$ is then multiplied, in the frequency domain, by the corresponding actuator model transfer function $A_i$ and by the optical response $O_i$ of the interferometer to motion of the mirror $i$. Finally, $h_{raw}$ is obtained by subtracting these reconstructed control contributions from the calibrated dark-fringe power.
Since the control signals are applied only up to a few hundred hertz, each of those subtraction terms is significant only below a few hundreds hertz, as illustrated in Fig.~\ref{fig:hrec_breakdown}.

Calibration lines injected into the control signals~$zC_i$ are used to monitor the optical response of the interferometer and to adjust every 4~s the time-dependent parameters of the optical response models used in the reconstruction. These parameters evolve slowly, following slow changes in the interferometer, for instance alignment drifts.\\
 

Once $h_{raw}(f)$ has been reconstructed, an operation referred to as {\it unbiasing} is applied. This is a new functionality developed for the O4 observing run. 
The frequency-dependent bias of $h_{raw}$, computed offline as described in section~\ref{sec:hrecerrors_weeklyO4}, is corrected in the frequency domain to obtain the unbiased strain data $h_{unbias}(f)$. When applied online, the correction introduces no additional latency.

This bias correction is updated manually whenever needed. However, it remained very stable throughout the O4 observing run, requiring only a few updates following major detector modifications, such as the replacement of the WE mirror or of the output photodiodes.\\

Finally, noise subtraction is applied. 
First, the calibration lines permanently injected (see Table~\ref{tab:callines}) to monitor bias and uncertainties are removed. 
Then, a broadband linear noise subtraction is applied to obtain the clean detector strain. This broadband linear noise subtraction has been improved with respect to the one used during O3: the new implementation accounts for possible correlations between different noise witness channels, as described in section~\ref{sec:NoiseSubtraction}.

All those intermediate and final $h(f)$ detector strain are then submitted to an inverse FFT producing the raw time seris $h_{raw}(t)$, the unbiased time series $h_{unbias}(t)$ and the final clean time series $h(t)$ (named V1:Hrec\_hoft\_16384Hz). This latter time series is distributed through the LVK network to low-latency and offline data analysis pipelines. Providing an online strain time series with negligible bias is a new feature of the O4 run. This is important because the strain data are used by the low-latency searches for compact binary coalescence and the match with the GW signals templates used in such searches would be slightly reduced in the presence of a bias.

\begin{figure}[!ht]
    \centering
	\includegraphics[trim={0 0cm 0 0cm},clip,scale=0.45]{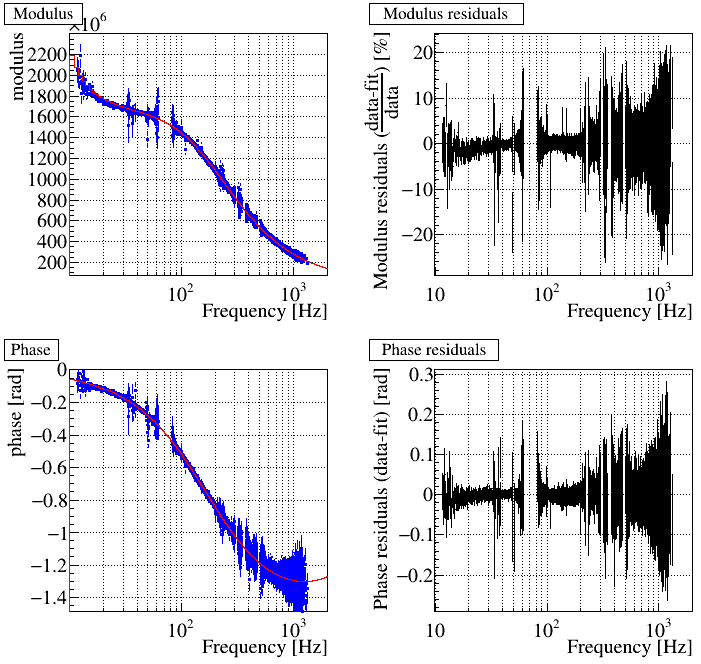} 
    \caption{Example of an optical response measurement performed during O4b using the NE mirror. The red curve shows a fit based on a model with a high-frequency single pole and a low-frequency double pole accounting the optical-spring. The right panel shows the residual between the measurement and the fit. In this measurement, the effective cavity pole frequency was found around 170~Hz, while the low frequency optical-spring double pole was located near 5~Hz. 
    }
    \label{fig:ORmeasurement}
\end{figure}

\subsection{Monitoring of time varying parameters: optical gains and double cavity pole}
\label{sec:hrecOR}

The Advanced Virgo+ detector includes both a Power Recycling mirror and a Signal Recycling mirror, each of which forms a coupled optical cavity with the interferometer. This modifies the standard optical response of a Fabry-Perot Michelson interferometer~\cite{bib:2026_DetectorPaper}. The two parameters describing this optical response, namely the optical gain and the double cavity pole frequency, were first estimated before the start of O4b and then checked periodically with broadband noise injections on NE and WE mirrors, while the fluctuations of those parameters were monitored online in the $h(t)$ reconstruction process.

The optical response model expected from the measurements made with broadband noise injections contains a single pole around 180~Hz and an increase of the modulus at very low frequency because of the impact of the optical spring double pole (which was maintained nevertheless below a few hertz by controlling the SR mirror position and alignment). Figure~\ref{fig:ORmeasurement} shows one of the optical response measurements done by injecting through the NE mirror actuator a broadband noise covering the 10-1000~Hz frequency band.\\

To monitor the time varying parameters of the optical response model used in the $h(t)$ reconstruction, permanent calibration lines (listed in Table~\ref{tab:callines}), are injected as sinusoidal excitations in the control signals  $zC_i$ of the WE, NE and BS before they are sent to the corresponding electromagnetic actuators~\cite{Virgocalib_2022}.

\begin{table}[tbh!]
\centering
\begin{tabular}{|c|c|c|c|}
\hline 
Injection source & Line frequency &  Line SNR in DARM\\ 
NE mirror actuator &  365.5 Hz  &  64  \\
WE mirror actuator &  363.5 Hz  &  60  \\
BS mirror actuator &  361.5 Hz  &  75  \\
\hline
NE mirror actuator &  71.5 Hz  &  38  \\
WE mirror actuator &  69.5 Hz  &  45  \\
BS mirror actuator &  67.5 Hz  &  60  \\
\hline
\end{tabular}
\caption{Permanent calibration lines injected during~O4 to monitor the optical gain (around 70~Hz) and the double cavity pole (around 360~Hz). The lines signal-to-noise ratios were estimated using the data acquired at GPS time 1441932000 (15 September 2025).
}
\label{tab:callines}
\end{table}

\begin{figure}[tbh!]
    \centering
	\includegraphics[trim={0 0cm 0 0cm},clip,scale=0.55]{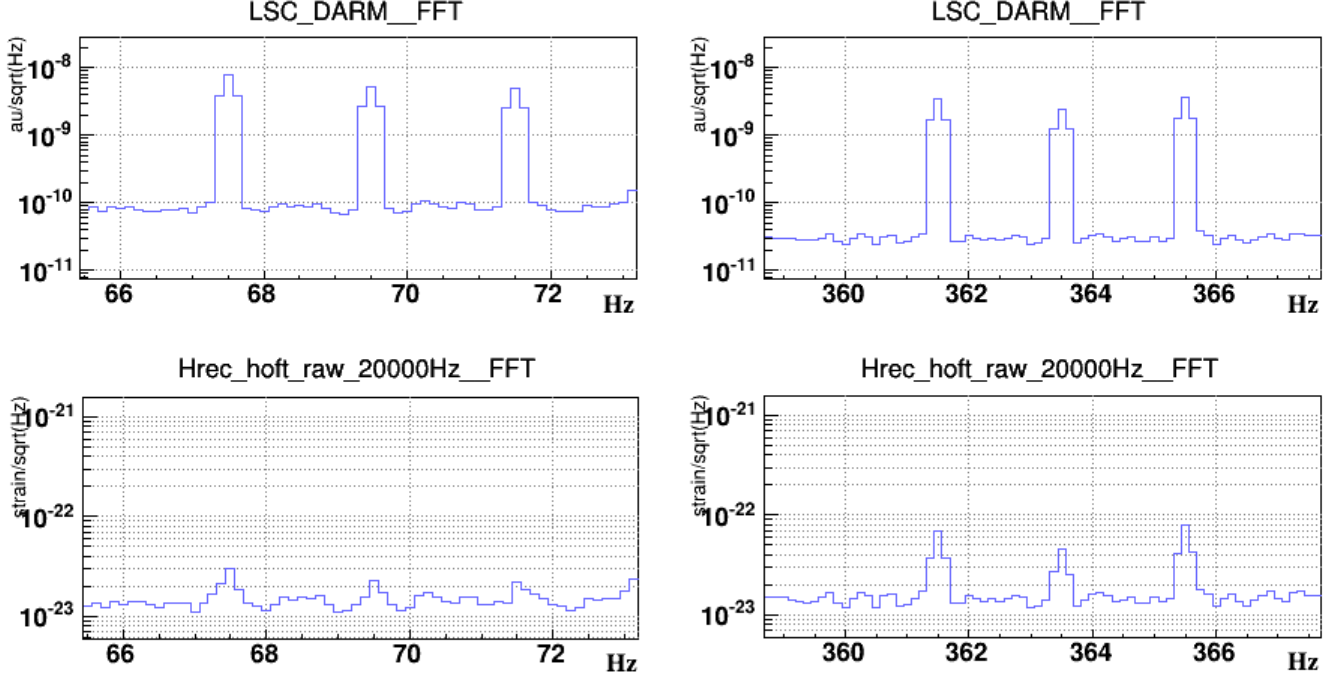} 
    \caption{Calibration lines visible in the spectrum of in the output dark fringe signal $\mathcal{P}_{DC}$  (upper plots) and in the raw reconstructed strain $h_{raw}$ (lower plots). The left plots show the lines around 70~Hz and the right plots shown the lines around 360~Hz. All spectra were computed using data acquired at GPS time 1441932000 (15 September 2025).
    The calibration lines are strongly suppressed in the reconstructed strain channel, as expected,
    although not completely removed because of residual inaccuracies in the actuator and optical-response models used by the reconstruction algorithm.
    }
    \label{fig:callinesO4}
\end{figure}

\newpage

A first set of lines, injected around 360~Hz, is used to monitor the double cavity pole. A second set of lines, around 70~Hz, is used to monitor the optical gains. These lines are injected with a signal-to-noise ratio of the order of~50 above the detector noise floor. Since they are included in the control signals $zC_i$, they are naturally removed by the $h(t)$ reconstruction algorithm.

\begin{figure}[tbh!]
    \centering
       \begin{minipage}[t]{0.8\textwidth}
	   \hspace{0.4cm}    \includegraphics[trim={0 0cm 0 0cm},clip,scale=0.45]{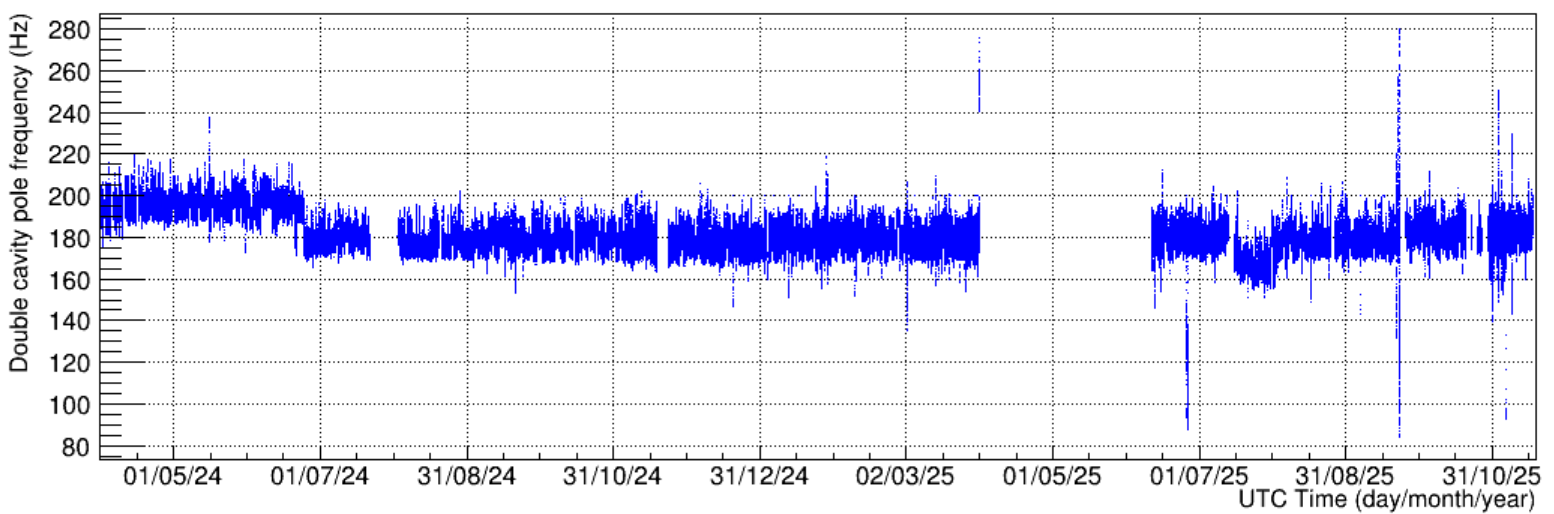}
    \end{minipage}
    \begin{minipage}[t]{0.8\textwidth}
        \hspace{0.4cm} 
	    \includegraphics[trim={0 0cm 0 0cm},clip,scale=0.45]{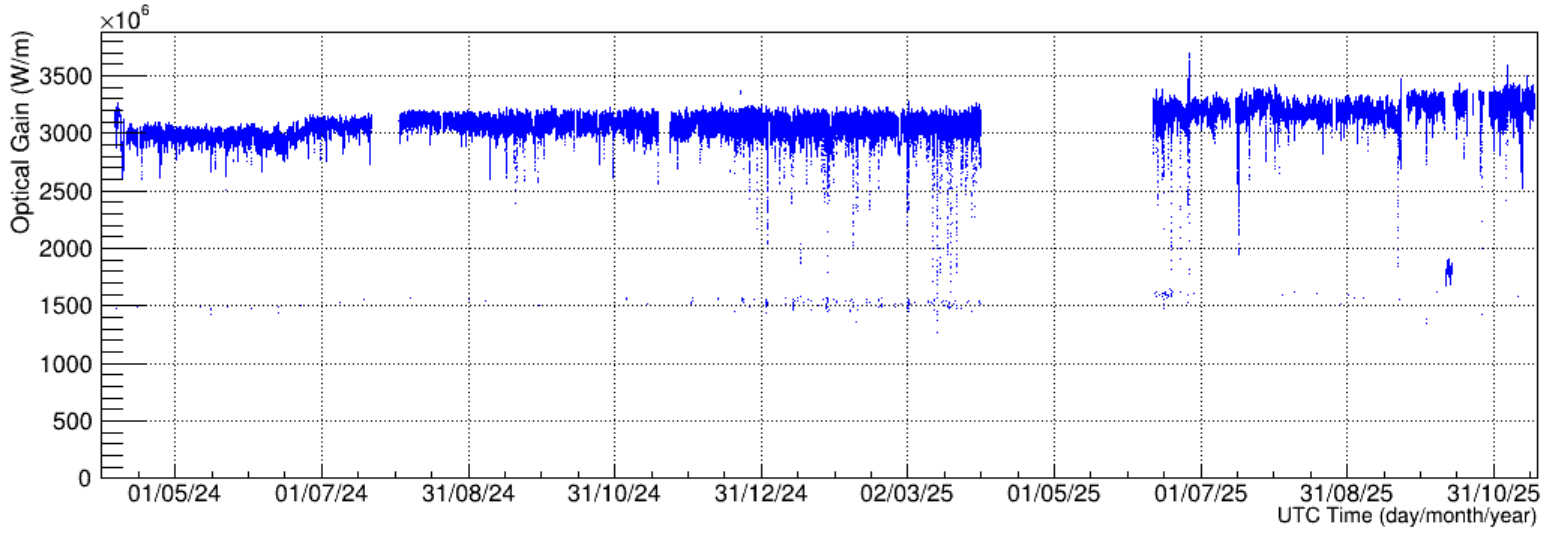}
    \end{minipage}
    \caption{Evolution of the double cavity pole frequency (Hz) and optical gain (W/m) of the average interferometer optical response throughout the O4b and O4c runs, as estimated by the detector strain reconstruction algorithm . 
    Both parameters exhibit good overall stability, although several features can be highlighted.
    On 24 June 2024, the pole frequency decreased from 195~Hz to 180~Hz while the optical gain increased. This change resulted from the choice of a better interferometer configuration, with an optimized misalignment of the SR mirror. 
    More generally, the optical gain was slightly larger after the April-June 2025 commissioning break.
    From 11 October 2025 10h24 UTC to 14 October 2025 06h10 UTC, the optical gain decreased from $3.2\times 10^9$~W/m down to $1.8\times 10^9$~W/m.
    This variation was associated to a temporary change of the interferometer operating point: following an issue affecting one of the output photodiodes between 8 and 11~October, the differential arm-length  offset was reduced, resulting in approximately a factor of two reduction in the interferometer output power.
    }
    \label{fig:gain_pole_O4}
\end{figure}

Using these lines, the optical gains $G_i$ and the double cavity pole frequency $fp_i$ (used in the optical response model $O_i$ shown in Fig.~\ref{fig:hrecprinciple}) are updated at the pace of the $h(t)$ reconstruction, namely once every 4~s. 
Figure~\ref{fig:callinesO4} shows the calibration lines visible in the spectrum of the photodiode output signal $\mathcal{P}_{CD}$, 
and their strong suppression in the reconstructed $h(t)$ spectrum. 
Figure~\ref{fig:gain_pole_O4} present the optical gain and the pole frequency estimated online by the $h(t)$ reconstruction process throughout O4b and O4c.

The low frequency features observed in optical response measurements, such as those shown in Fig.~\ref{fig:ORmeasurement}, are not modelled in the strain reconstruction algorithm. This may induce a small bias in the raw reconstructed strain. This bias is nevertheless corrected for through the unbiasing procedure, as described later. Hence, there were no significant impact of this approximation on the final strain data.

\subsection{Noise subtraction}
\label{sec:NoiseSubtraction}

Some noise may couple to the differential arm length (DARM) variations and not be fully cancelled by the detector control loops: it therefore adds extra noise to the reconstructed detector strain and degrades the sensitivity of the detector. 
When available and identified, noise-witness channels are used to estimate these extra noise contributions and mitigate the remaining noise in the reconstructed detector strain $h_{raw}(t)$.
The noise subtraction computation is performed in the frequency-domain and is done by the same process Hrec that takes care of the $h(t)$ reconstruction.\\

For each noise-witness channel, a transfer function (TF) between $h_{raw}$ and the witness channel is computed in a specific frequency band to quantify the noise contribution that should be subtracted. 
During O4, the noise subtraction used a new method based on~\cite{driggers2019improving,bib:ligo_noisesub}, that allows the linear subtraction of multiple noise even in the case of partially correlated noise. More details on the noise subtraction techniques used during O3 and modified for O4 can be found in~\cite{bib:TDSHrec}.
During O4, in order to monitor slow variations of the noise coupling in the interferometer, the TFs were updated every 240~s and used to subtract the associated noise over the following 240~s.

\begin{figure}[tbh!]
\centering
	\includegraphics[trim={0 0cm 0 0cm},clip,scale=0.55]{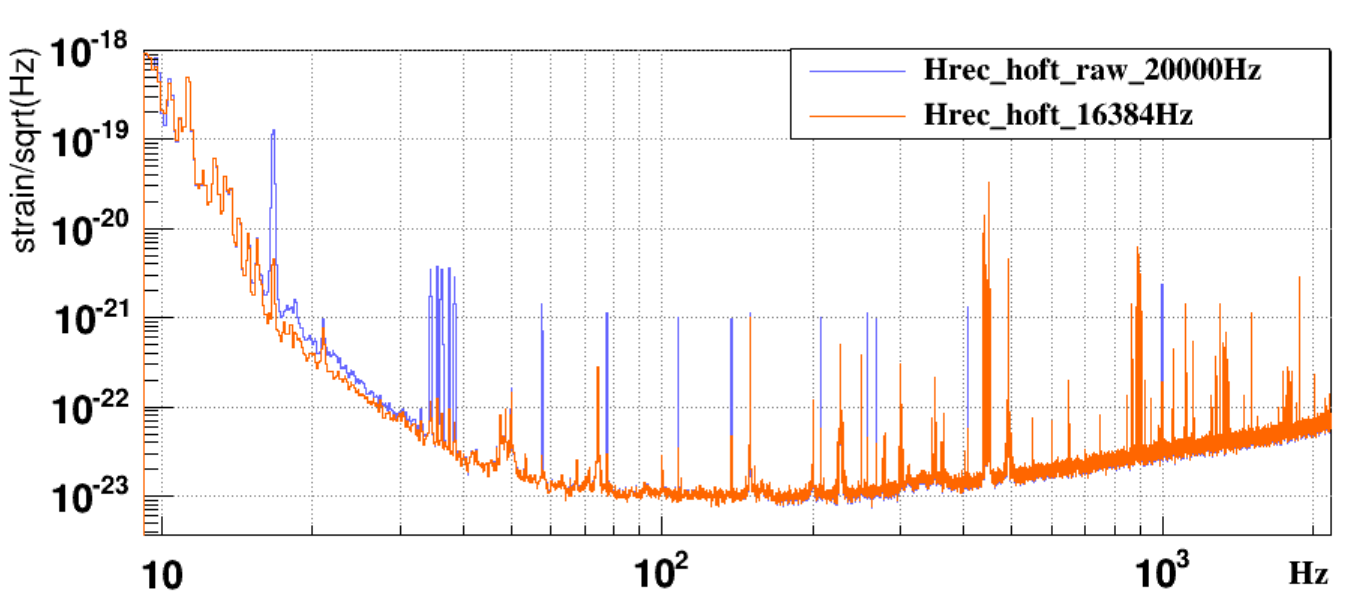}
    \caption{Typical Virgo O4 sensitivity curve before (blue) and after (red) noise subtraction: amplitude spectral densities of $h_{raw}(t)$ and $h(t)$ on 20 April 2024, 11h49m42s~UTC (GPS=1397649000). The BNS range associated to this sensitivity curve is 56~Mpc.
    }
    \label{fig:O4sensitivity}
\end{figure}

\noindent
Two witness channels identified before the O4 observing run were used to subtract the associated noise:
\begin{itemize}
  \item Michelson control noise: The motion of the BS mirror creates a differential signal between the two interferometer arms. Compared to the arm cavity mirrors the impact of this motion is reduced by the optical gain of the Fabry-Perot cavities. However, the Michelson control noise is much higher than the differential arm length control noise; therefore, it yields a non-negligible contribution to the overall interferometer noise. To subtract this noise contribution, the corresponding control signal  (called LSC\_MICH) is used as witness channel in the frequency band 8-200~Hz.
  \item Signal Recycling control noise: The motion of the SR mirror creates a differential signal between the two interferometer arms. Thus, any noise present in the control loop dedicated to the SR longitudinal motion can be introduced in the differential signal. It contributes mainly below 40~Hz. To subtract this noise contribution, the corresponding control signal (called LSC\_SRCL) is used as a witness channel in the frequency band 8-40~Hz.
\end{itemize}

During O4, the witness channels LSC\_MICH and LSC\_SRCL were partially coherent with each other in the band 8 to 40~Hz. The new noise subtraction method allowed to disentangle the contributions of those two channels.

\begin{figure}[tbh!]
\centering
	\includegraphics[trim={0 0cm 0 0cm},clip,scale=0.47]{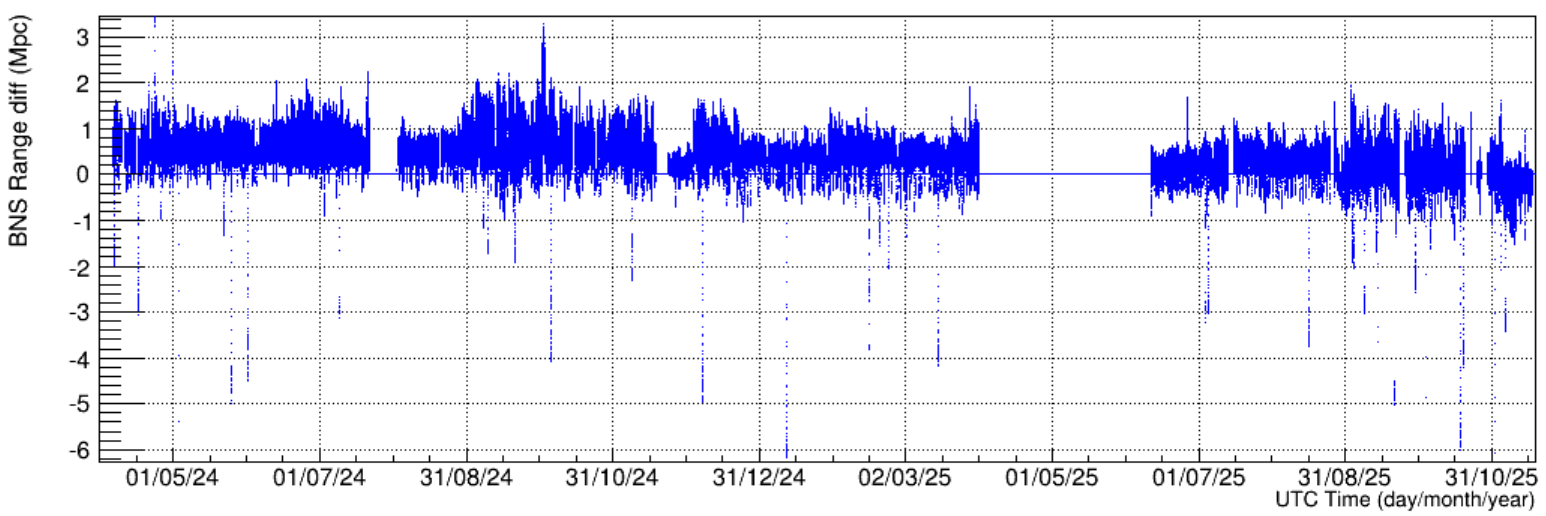}
    \caption{Evolution of the BNS range improvement due to noise subtraction. This is the difference between the BNS range computed with $h_{raw}$ and the BNS range computed with the final $h(t)$ after noise subtraction. The few time periods where this difference is negative correspond to large glitches in the data which perturbed the transfer function used in the noise subtraction procedure. A protection against such effect has been implemented in Hrec but not yet tested.
    }
    \label{fig:bnsrangediff}
\end{figure}

\noindent
The noise subtraction applied during O4, allowed a small improvement of the overall detector sensitivity, especially in the 8-40~Hz frequency band, as shown on the Fig.~\ref{fig:O4sensitivity}.
It corresponds to an improvement of 0.5 to 1~Mpc on the binary neutron star (BNS) range~\footnote{The BNS range is defined as the distance, averaged over all sky directions and source orientations, at which a binary neutron star merger is seen by the interferometer with a signal-to-noise ratio of~8.} \cite{bib:GWTC5intro}. Figure~\ref{fig:bnsrangediff} shows the evolution of this improvement throughout O4b and O4c.

\subsection{Performance of the online detector strain reconstruction during O4}
\label{sec:Hrecperf}

\begin{figure}[tbh!]
\centering
    \begin{minipage}[t]{0.8\textwidth}
	   \hspace{0.4cm}    \includegraphics[trim={0 0cm 0 0cm},clip,scale=0.45]{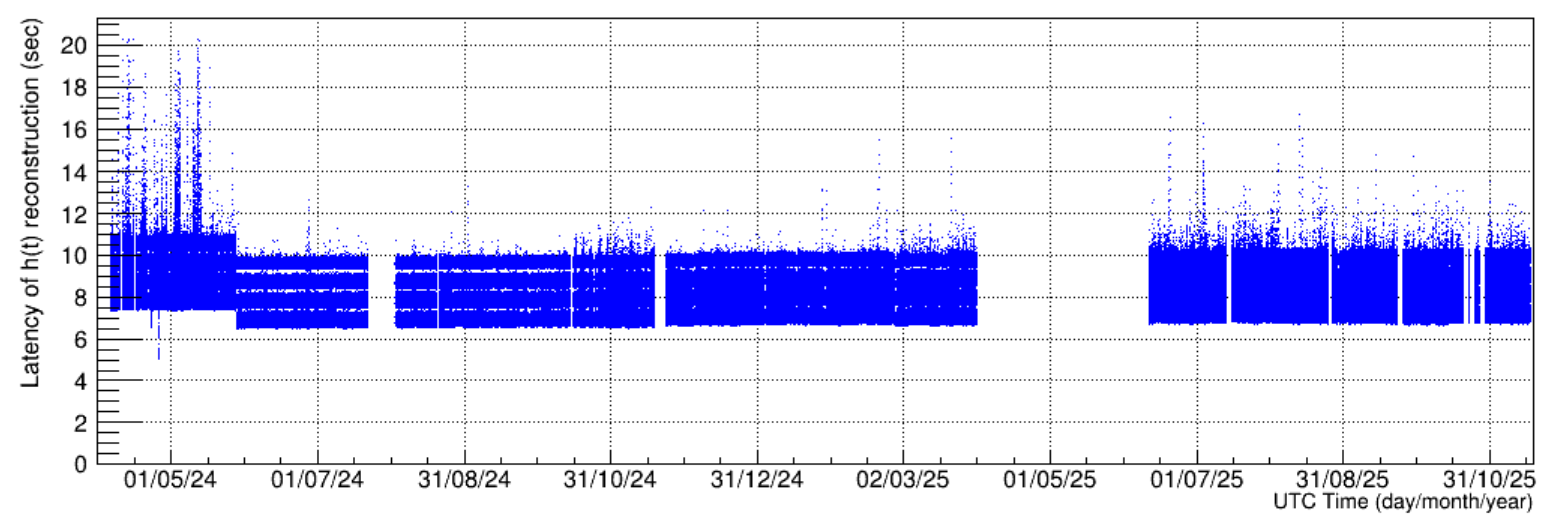} 
    \end{minipage}
    \begin{minipage}[t]{0.8\textwidth}
        \hspace{0.4cm} 
	    \includegraphics[trim={0 0cm 0 0cm},clip,scale=0.45]{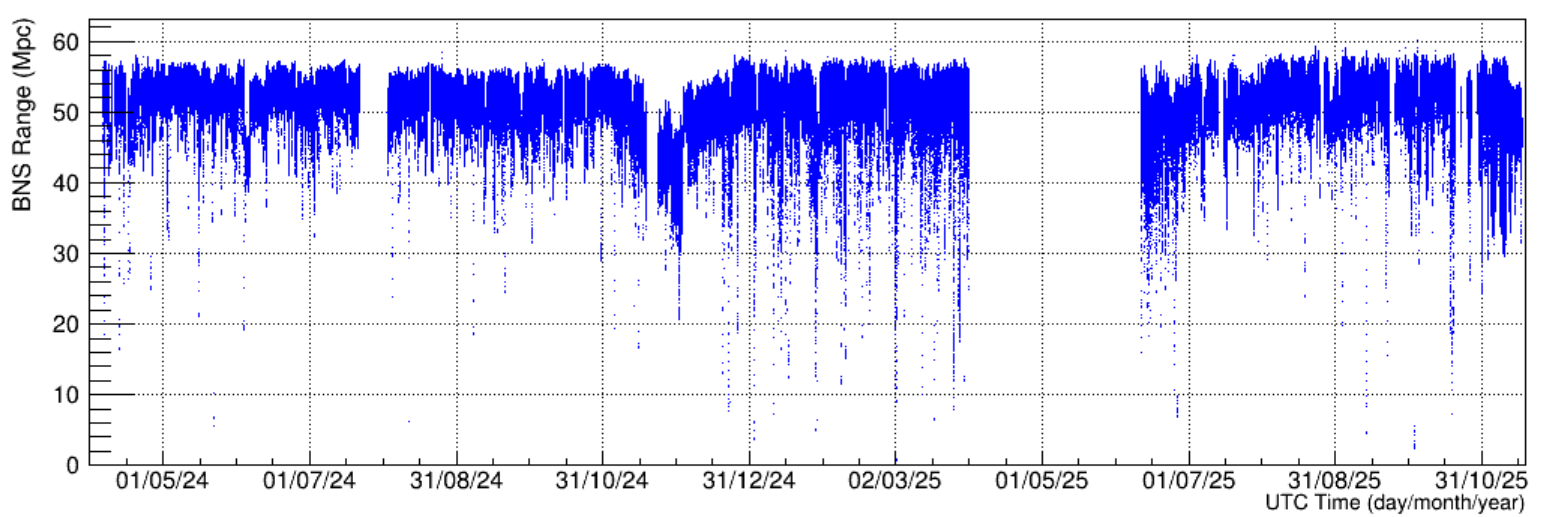}
    \end{minipage}
    \caption{
    Top: latency (in s) of the online Virgo detector strain reconstruction over the O4 run. 
    The four horizontal bands are linked to the way the reconstruction algorithm works, taking input data by chunks of 4~seconds. 
    On 28 May 2024, an improvement on the data collection chain allowed to gain one second of latency for the availability of the time series at the input of the reconstruction processing.\\
    Bottom: Virgo BNS range (in Mpc) over the O4 run. 
    Points are shown only when data quality is good and data are within observation periods of the interferometer).
    Drops of the BNS range down to 45 or 40~Mpc are due to somewhat periodic glitches that affected Virgo during O4b and O4c
    Drops down to 30 or 20~Mpc are due to larger glitches often linked to photodiodes saturation.
    }
    \label{fig:Hrecperf}
\end{figure}

The online $h(t)$ reconstruction process
was active for more than 99.9\% of the O4b+O4c time period. It delivered the $h(t)$ time series with a mean total latency of about 9.3 seconds. The upper plot of Fig.~\ref{fig:Hrecperf} shows the evolution of the latency of the strain time series delivery between April~2024 and November~2025, which remained quite stable after 28 May 2024.
The lower plot of Fig.~\ref{fig:Hrecperf} shows, over the same time period, the evolution of the BNS range, after noise subtraction.\\

Another figure of merit is the bias of the strain reconstruction.
As described earlier, the bias of the raw detector strain $h_{raw}(t)$ is corrected online to provide an unbiased detector strain time series $h(t)$. This bias is monitored continuously, as described in~\ref{sec:bias_permmoni}.
Fig~\ref{fig:Hrechinj} shows that the bias modulus of $h(t)$
evolved during the O4b+O4c time period, but was quite stable around 1. It shows that the online unbiasing performed properly throughout the run.



\section{Estimation of h(t) bias and uncertainties}
\label{sec:hrecerrors}

The $h(t)$ time series, described in the previous section, is the main Virgo data product for GW data analysis. However, the knowledge of its bias and uncertainty is also required, in particular for the estimation of the physical parameters of detected sources~\cite{bib:GWTC4methods}. During~O4, this information was periodically stored in the frame files together with the $h(t)$ time series.

Several types of measurements were performed before and during the run to quantify the bias and uncertainty of $h(t)$. For most of them, calibration lines were injected to drive the NE and WE end mirrors with a known strain-equivalent signal $h_{inj}$, which was then compared to the reconstructed strain $h_{raw}$.

\begin{itemize}
    \item Injections of 27~lines between about 10~Hz and 1500~Hz were run over a few minutes every week, with a high signal-to-noise ratio. Such data were used to estimate the bias of the reconstructed $h(t)$ time series, either the bias of the raw $h_{raw}(t)$ time series (to be corrected for in the reconstruction as described in section~\ref{sec:hrec}), or the residual bias of the final (unbiased) $h(t)$ time series.
    \item Broadband noise injections were also periodically performed through the mirror actuators to provide a snapshot of the bias over the full detection frequency band. These measurements were mainly used to look for possible frequency regions exhibiting a bias larger than what could be detected using only the 27 calibration lines.
    \item 11 permanent lines were injected through the NE mirror electromagnetic actuator, with a lower signal-to-noise ratio, between 16.7~Hz and 997.5~Hz, as listed in Table~\ref{tab:permlines}. They were used to monitor the stability of the $h(t)$ bias as a function of time. The observed variations were then used to estimate the frequency-dependent systematic uncertainties of the $h(t)$ time series. 
\end{itemize}

\begin{table}[!ht]
\centering
\begin{tabular}{|c|c|c|c|}
\hline 
Injection source & Line frequency &  Line SNR in DARM\\ 
\hline
NE mirror actuator &  16.7 Hz   &  8  \\
NE mirror actuator &  37.5 Hz   &  80  \\
NE mirror actuator &  57.5 Hz   &  60  \\
NE mirror actuator &  77.5 Hz   &  60  \\
NE mirror actuator &  107.5 Hz  &  60  \\
NE mirror actuator &  137.5 Hz  &  60  \\
NE mirror actuator &  207.5 Hz  &  60  \\
NE mirror actuator &  257.5 Hz  &  65  \\
NE mirror actuator &  267.5 Hz  &  50  \\
NE mirror actuator &  407.5 Hz  &  50  \\
NE mirror actuator &  997.5 Hz  &  45  \\
\hline
WE mirror actuator &  35.5 Hz   &  65  \\
WE mirror actuator &  996.5 Hz  &  55  \\
\hline
NE Pcal &  34.5 Hz   &  55  \\
NE Pcal &  999.5 Hz  &  8  \\
WE Pcal &  38.5 Hz   &  52  \\
WE Pcal &  994.5 Hz  &  7  \\
\hline
\end{tabular}
\caption{Frequencies of the lines permanently injected on NE and WE mirrors (using either the electromagnetic actuators or the Pcal actuators).
The lines injected on NE mirror actuator (top) are used to monitor the bias and to estimate the uncertainties on $h(t)$.
The other lines are used to monitor the stability of the cross-calibration of the different actuators.
The signal-to-noise ratio (SNR) of the lines has been estimated using 10~s Fast-Fourier Transforms (FFTs) at a given time during O4b. They may vary by a few percent depending on the variations of the noise level in the running interferometer. Low SNR at 7 and 8 are due to the limitation of the PCal to inject high amplitude signal at high frequency.}
\label{tab:permlines}
\end{table}

\subsection{Extraction of h(t) bias from the measurements}
Before presenting the bias monitoring results obtained with the different types of measurements, we first describe the method used to extract the bias information from the various calibration injections. This is done by computing the ratio $h_{raw}/h_{inj}$, where $h_{inj}$ is the injected signal computed using the actuators models and where $h_{raw}$ is the raw reconstructed strain (before unbiasing), as described in section~\ref{sec:hrec}.

The injected strain $h_{inj}$ is computed using the amplitude of the signal of the calibration lines injections and the actuators models estimated by the calibration procedure~\cite{Virgocalib_2022}:
\begin{equation}
\label{eqn:Hinj}
  h_{inj} = N_i \times \frac{A_i}{L_0} \times e^{-2j\pi f \tau} . \frac{TF_{true}}{TF_{pole}}
\end{equation}
where $N_i$ is the excitation channel used to inject the calibration lines through the actuator $i$ and $A_i$ is the actuator transfer function. The last factor in Eq.~(\ref{eqn:Hinj}) accounts for two additional effects:
\begin{enumerate}
    \item The detector response to the motion of an end mirror exhibits an additional delay $\tau=10\,\mu s$ compared to the response to a passing GW~\cite{bib:Rakhmanov_2008_shortwavelengthapprox,bib:2015_CalibApproximations}. This delay is taken into account when computing $h_{inj}$.
    \item The simple pole cavity approximation used in the optical response model for $h(t)$ reconstruction is compensated by the long-wavelength approximation in the data analyses when computing the antenna pattern~\cite{bib:2015_CalibApproximations}. 
    Consequently, $h_{raw}$ contains a bias due to the simple pole approximation, that should not be corrected, since it is compensated later in the data analysis chain. The ratio $TF_{true}/TF_{pole}$ is therefore included to apply the same bias to the estimated $h_{inj}$. This bias is lower than 0.5\% in the modulus up to 2~kHz, and corresponds to a timing offset of approximately -13\,\mus\ in the phase (see~\cite{bib:2015_CalibApproximations}).
 \end{enumerate}
 
As $h_{raw}$ and $h_{inj}$ share the same bias from the simple pole approximation, the ratio $h_{raw}/h_{inj}$ computed in this way is expected to be equal to 1 in modulus and 0 in phase. 
Small deviations from these values provide the frequency-dependent bias $B_{raw} = h_{raw}/h_{inj}$ of the reconstructed strain time series at a given time.

When computed using the $h_{raw}$ time series, this bias $B_{raw}$ is then provided as input to the $h(t)$ reconstruction process to correct the bias in $h_{raw}$ and obtain $h_{unbias}$, as described in section~\ref{sec:hrec}. 
For the online reconstruction, $B_{raw}$ was estimated from past data and subsequently used in the reconstruction process. The bias correction was kept fixed over long periods and updated only a few times during O4, after changes due to hardware interventions.
Since the correction is fixed and derived from past measurements, a residual bias $B_{res}=h_{unbias}/h_{inj}$ can be present on the final $h(t)$ time series. This residual bias was continuously monitored using the set of injections at the 11 different frequencies mentioned above to determine whether an update of the online bias correction was required.\\

\subsection{Estimation of the bias of the raw detector strain}
\label{sec:hrecerrors_weeklyO4}

\begin{figure}[tbh!]
  \centering
  \includegraphics[width = 0.7\textwidth]{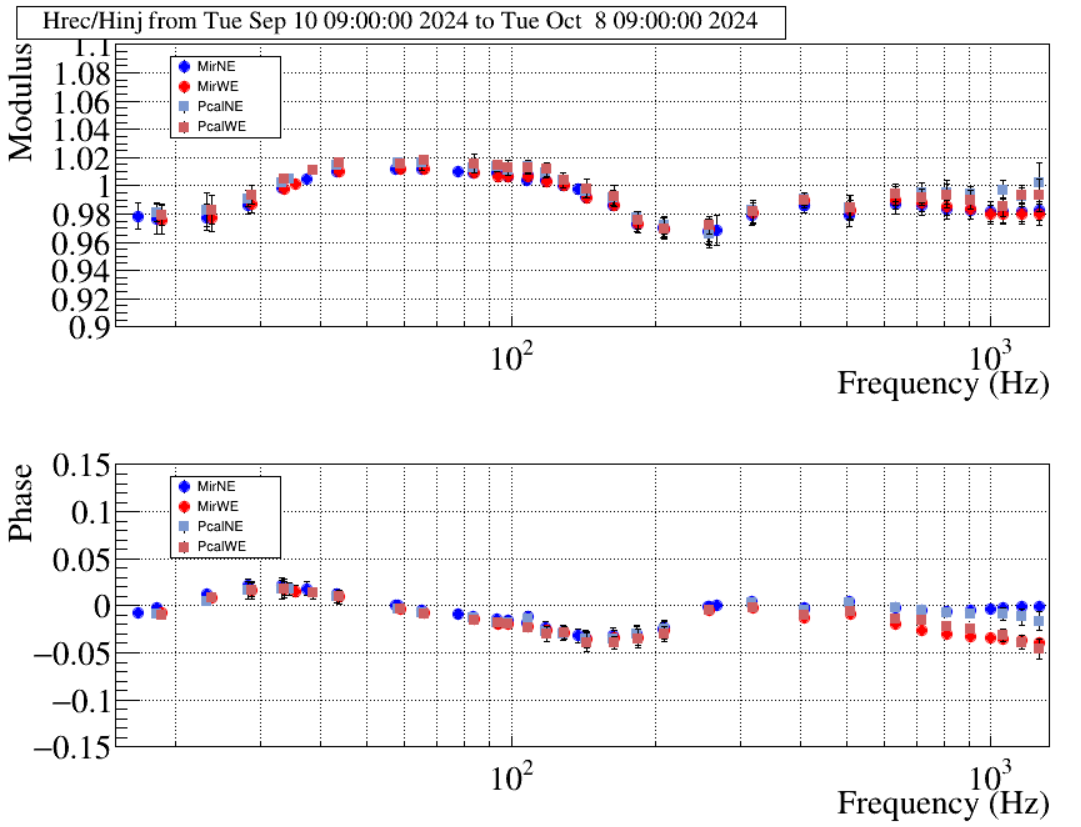}
  \caption{Modulus and phase of the $h_{raw}(t)$ bias estimated from weekly injections between 10 September 2024 and 8 October 2024, before the application of the bias correction.
  }
  \label{fig:O4_weeklylines}
\end{figure}

Figure~\ref{fig:O4_weeklylines} shows measurements of $B_{raw}=h_{raw}/h_{inj}$
obtained from the dedicated injections of 27~lines and averaged during the time period from 10~September 2024 to 8~October 2024. 
The values of $B_{raw}$ are close to~1 in modulus and close to~0 in phase, as expected. However, a systematic frequency-dependent bias is observed and is consistent across the four different actuators used for the injections. This bias varies with frequency and reaches a maximum of about 4\% in modulus and 50~mrad in phase.

\begin{figure}[ht]
  \centering
  \includegraphics[width = 0.7\textwidth]{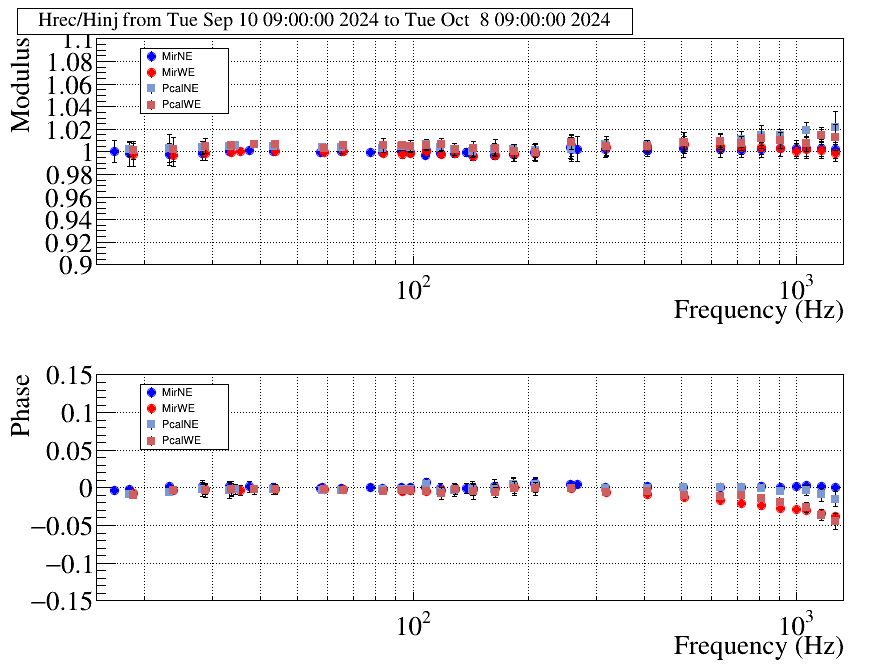}
  \caption{\label{fig:CheckHrec_unbias} Modulus and phase of the $h_{unbias}(t)$ residual bias estimated from weekly injections between 10 September 2024 and 8 October 2024.
  }
\end{figure}

The bias measured around 36~Hz by the four actuators is also independently monitored and confirmed by the Newtonian calibrators.
The agreement between measurements obtained with different actuators rules out errors in the estimation of the injected signal $h_{inj}$ and confirms the presence of an intrinsic bias in the $h_{raw}(t)$ time series itself. 

At high frequency, however, a discrepancy of about 25~mrad around 1~kHz is observed between the phases of $B_{raw}$ measured using the NE and WE actuators. This corresponds to a timing difference of approximately 4\,\mus\ and indicates a difference in the timing calibration of the actuators, consistent with the 3\,\mus\ uncertainty estimated for the mirror actuators in Table~\ref{tab:MirUncertainties}.

The overall bias of the $h(t)$ time series is computed using only the injections done through the NE electromagnetic actuator. A method was developed to interpolate the bias between the measured frequencies, with a frequency resolution of 0.125~Hz~\cite{PhdGrimaud, bib:calib_Amaldi}. 
When computed on the $h_{raw}$ time series, this bias information $B_{raw}$ is used to derive the bias correction to be applied in the reconstruction processing to get the final "unbiased" detector strain time series.

\subsection{Estimation of the residual bias of the final detector strain} \label{sec:hrec:bias}

\subsubsection{Estimation of the residual bias with the calibration lines} \label{sec:hrec:bias:lines}
The same analysis applied to the unbiased time series $h_{unbias}(t)$ is shown in Fig.~\ref{fig:CheckHrec_unbias}.
In this case, the residual bias is indeed negligible, remaining
below 0.5\% in modulus and 10~mrad in phase. 
Since the bias correction is derived from the NE mirror actuator injections, the residual bias measured with those injections is flat in both modulus and phase. In contrast, a residual phase bias corresponding to approximately 4\,\mus\ bias is observed when using the WE mirror injections.

This residual bias information $B_{res}$, computed on the $h_{unbias}$ time series, is stored and distributed together with the $h(t)$ time series as an additional data product for offline GW analyses.

\subsubsection{Verification of the residual bias with broadband injections} \label{sec:hrec:broadband}
\label{sec:broadband_inj}


\begin{figure}[tbh!]
  \begin{center}
    \subfigure[Full frequency band] {
      \label{fig:CheckHrec_MIR_NE_50Hz}
    \includegraphics[angle=0,width=0.68\linewidth]{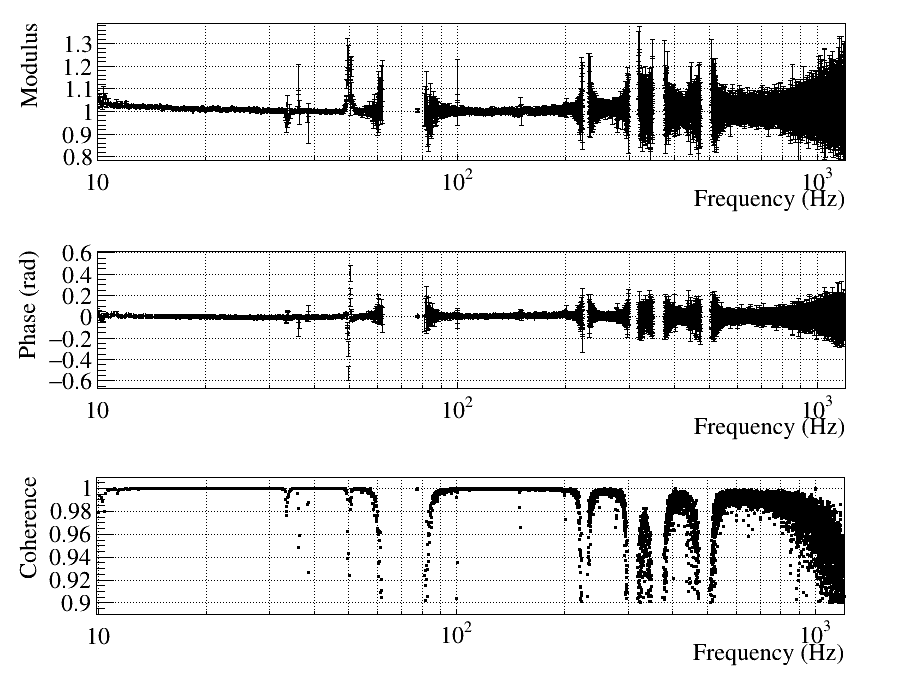}
      }
    \subfigure[Zoom around 50~Hz] {
      \label{fig:CheckHrec_MIR_NE_50Hz_zoomed}
\includegraphics[angle=0,width=0.68\linewidth]{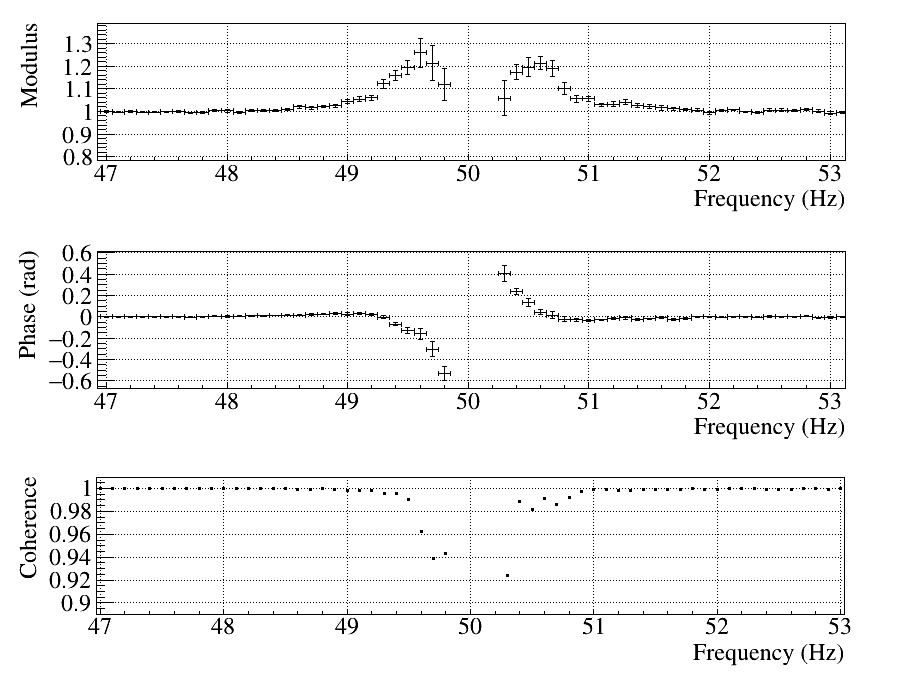}
      }
    \caption{Broadband measurement of transfer function $h_{unbias}/h_{inj}$ performed using the NE actuator on 14 January 2025. 
    Top: modulus, phase and coherence values of the transfer function. 
    Bottom: zoom around 50~Hz.
    The FFTs are computed over 10~s time windows and only points with coherence above 0.9 are retained.
    }
  \end{center}
\end{figure}

Figure~\ref{fig:CheckHrec_MIR_NE_50Hz} shows the ratio $h_{unbias}/h_{inj}$ obtained from a broadband noise injection performed through the NE mirror electromagnetic actuator, during O4: the bias is close to 1 in modulus and 0 in phase, as expected. 
Overall, the measured bias agrees with the estimates obtained from the 27~calibration lines, except around 50~Hz (the mains frequency in Europe) and 150~Hz (its second harmonic). In a few frequency bins around 50~Hz, deviations as large as 20\%-30\% in amplitude and 600~mrad in phase, are observed, as shown in Fig.~\ref{fig:CheckHrec_MIR_NE_50Hz_zoomed}.

The broadband injections therefore confirm that the bias $B_{res}$ estimated with the 27~lines is correct, except in the vicinity of 50~Hz and 150~Hz where the bias is larger. This excess bias is not corrected but instead incorporated in the systematic uncertainty budget of $h(t)$, as discussed in the next section. 
These frequency bands are strongly contaminated by mains-related noise and its harmonics, where the detector sensitivity is degraded,
so the impact on the overall scientific outcomes is minor.

\subsection{Estimation of the h(t) uncertainties using the permanent lines}
\label{sec:bias_permmoni}

\begin{figure}[tbh!]
\centering
    \begin{minipage}[t]{0.8\textwidth}
        \hspace{0.4cm} 
	    \includegraphics[trim={0 0cm 0 0cm},clip,scale=0.5]{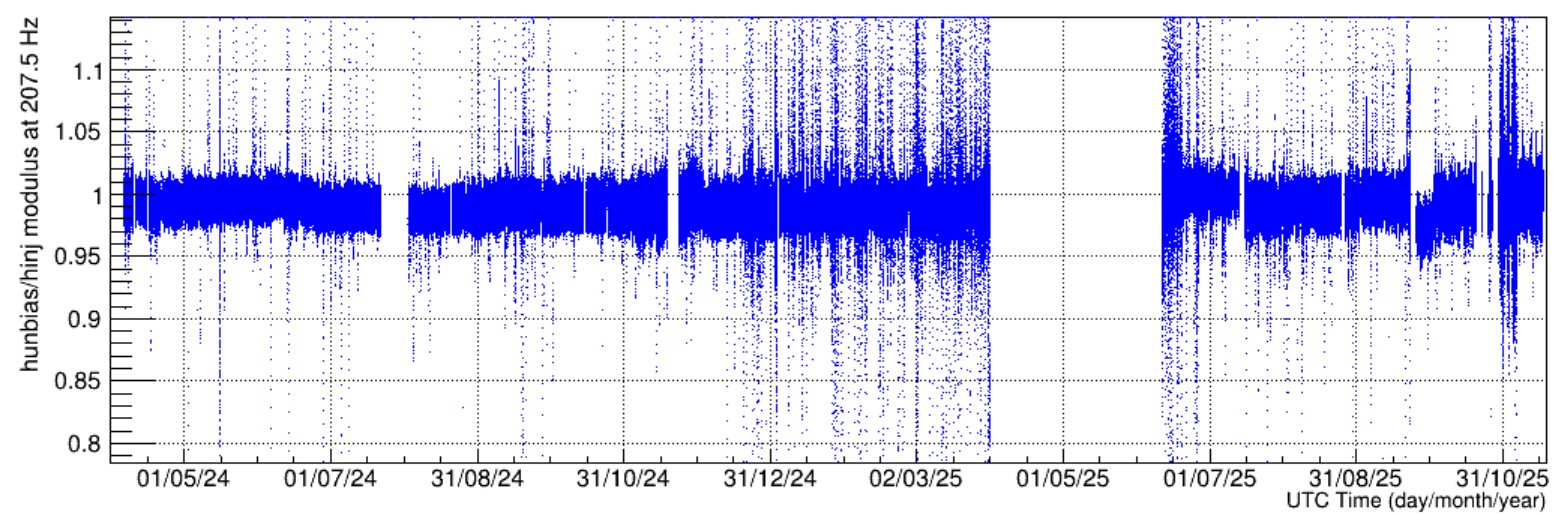}
    \end{minipage}
    \caption{
    Evolution of the residual bias of the detector strain data monitored at 207.5~Hz.
    The modulus average value remains very close to~1, confirming that the initial raw bias (monitored between -4\% and -10\% over O4) is properly corrected.
    The dominant variations are of the order of $\pm 2\%$.
    Large excursions are caused by transient glitches whose bandwidth overlaps the 207.5~Hz frequency and which temporarily perturb the monitoring.
    }
    \label{fig:Hrechinj}
\end{figure}

The 11 permanent lines described at the beginning of this Section and listed in Table~\ref{tab:permlines} were continuously injected through the NE mirror actuator during O4 and were used to estimate the
uncertainties on $h(t)$.
The modulus and phase of the $h_{unbias}/h_{inj}$ ratios at the injected
frequencies were computed online using a moving average of twelve 10~s long FFTs and were provided in the online data stream as channels sampled at 1~Hz~\cite{bib:TFMoni_doc,bib:2021_TFMoni_O3}. 
The $h_{inj}$ signals were estimated using the actuators models established at the beginning of~O4. 
Figure~\ref{fig:Hrechinj} shows, as an example, the modulus of $h_{unbias}/h_{inj}$ monitored at 207.5~Hz throughout the O4 run.
Such a monitoring is a test of possible variations of the bias of the reconstructed strain data, as well as a test of possible variations of the NE actuator response. In practice, all observed variations, at the level of $\pm 2\%$ in modulus, are included in the uncertainty budget of $h(t)$. This is consistent with the fact that the measured stability of the actuator response is significantly better than 2\%. It is also expected that the bias of $h(t)$ varies at the percent level, in particular because the slow monitoring of the optical response parameters is itself affected by uncertainties. The optical models are only approximations and the estimation of their parameters is statistically limited by the amplitude of the injected calibration lines listed in Table \ref{tab:callines}.

In practice, the distributions of the $h_{unbias}/h_{inj}$ time series at the 11 calibration frequencies are computed over periods of about one month. 
The distributions are then interpolated into a frequency-dependent distribution with a frequency resolution of 0.125~Hz~\cite{PhdGrimaud, bib:calib_Amaldi}. 
Then, the rms of the interpolated distributions is subsequently computed for each frequency bin.
Finally, to obtain the total uncertainty on the detector strain time series $h(t)$, 
the uncertainty on the signal $h_{inj}$, derived from the NE mirror actuator calibration, is added in quadrature.

\begin{figure}[bph!]
  \begin{center}
      \includegraphics[angle=0,width=0.7\linewidth]{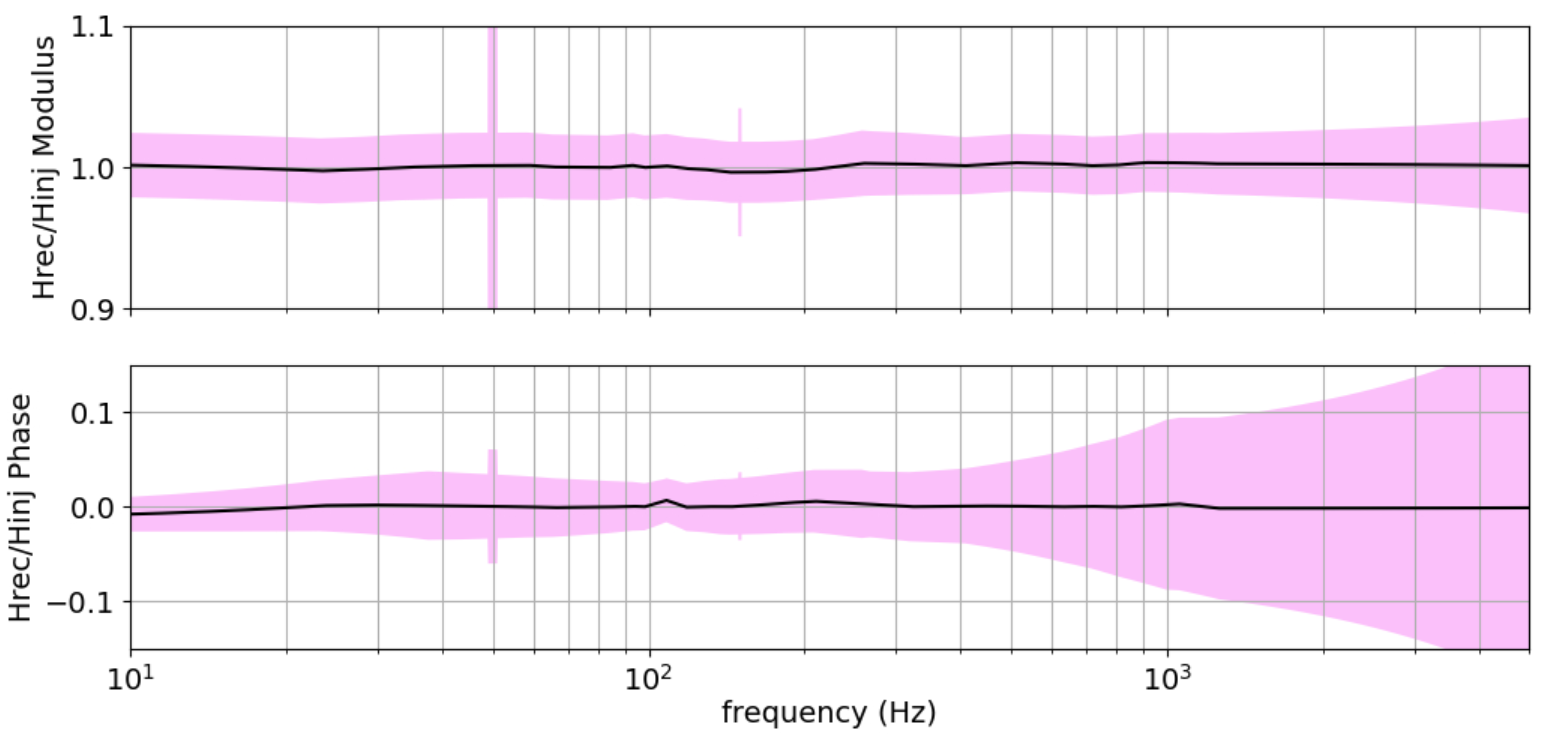}
    \caption{Residual bias and uncertainty computed from 
    $h_{unbias}/h_{inj}$ measurements for the period from 10~September to 8~October 2024. 
    The black curve corresponds to the residual bias of the $h(t)$ time series, while the pink shaded region represents the $\pm1\sigma$ uncertainty around the bias.
    }
    \label{fig:O4_uncertainty}  
  \end{center}
\end{figure}

At the beginning of O4b, and later when computing the $h(t)$ final uncertainties, the uncertainty on the NE actuator was estimated to be 0.78\% on the modulus (0.6\% from the Pcal calibration and 0.5\% from the transfer calibration between the Pcal and the mirror actuator), 5~mrad in phase and 5~\mus\ in timing.
These values slightly overestimate the uncertainties reported in this article since the most recent estimates are 0.77\% on the modulus and 3~mrad on the phase as summarized in Table~\ref{tab:MirUncertainties}. The timing uncertainties are estimated to be 1\,\mus\ for the interferometer photodiode readout chain and 3\,\mus\ for the mirror actuators. 

Figure~\ref{fig:O4_uncertainty} shows the residual bias and the associated frequency-dependent symmetric uncertainty, computed for the period from 10~September to 8~October 2024.
As expected, the residual bias is negligible.
The uncertainty is approximately 2.5\% in modulus and below 40~mrad in phase, then increasing at high frequency due to the timing uncertainty.

Such values are dominated by the temporal variations in the reconstructed strain rather than by uncertainties in the NE mirror actuator model. Consequently, the slight overestimation of the actuator calibration uncertainty has a negligible impact and mainly affects the phase uncertainty at high frequency through the timing contribution.\\

The uncertainties in the vicinity of 50~Hz (49-51~Hz) and 150~Hz (149-151~Hz) were increased to encompass the larger bias described in
section~\ref{sec:broadband_inj}, reaching 20\%/50~mrad and 4\%/20~mrad respectively.

It should be noted that there is a known inconsistency in this estimate. As can be seen on Fig.\ref{fig:CheckHrec_MIR_NE_50Hz_zoomed}, the phase uncertainty around 50~Hz should be significantly larger, of the order of 500~mrad. However, at these frequencies, the detector sensitivity is strongly contaminated by electrical mains noise and, therefore, the corresponding frequency bins can not contribute significantly
to any data analyses. In practice, the LVK parameter estimation analyses completely remove these two frequency bands from their analysis and therefore do not rely on the associated uncertainty estimates~\cite{bib:GWTC5_methods_orPEpaper}.

\section{Production of the Analysis Ready frame files}
\label{sec:ARframes}

The $h(t)$ reconstruction was running online during O4, and the $h(t)$ time series, as well as some auxiliary channels, were distributed with a latency of about 10~s down to the GW searches for public alert generation. The same data products were also stored on disk into the so-called aggregated frame files of 2000~s for offline use.

An additional set of files was produced later by the Virgo Calibration and Detector Characterization groups: the so-called Analysis Ready (AR) frame files.
These files are the ones used by the members of the LVK Collaboration to run most of offline analyses and extract the information published by the collaboration \cite{bib:GWTC5_methods_orPEpaper, bib:GWTC5}. These same files are also made publicly available on the GWOSC web site~\cite{bib:GWOSC,bib:GWOSCpaper}.\\

In this section, we provide a brief description of the content of these AR frame files. 
We then summarize how the files were produced and validated.
For full details about the O4b and O4c AR frames production and checks, see \cite{bib:ARframes_prod,bib:checks_ARframes_O4b,bib:checks_ARframes_O4c}.

\subsection{AR frame files content}
\label{sec:ARframecontent}
During O4, the Virgo data products were provided in frames~\cite{bib:FrameFormat} that contain the $h(t)$ time series sampled at 16384~Hz,
a so-called state vector describing the data quality sampled at 1~Hz,
and, only stored once every 1000~s, six vectors containing the information about the frequency-dependent bias and uncertainty.
The names of the channels are the following:

\begin{itemize}
    \item V1:Hrec\_hoft\_16384Hz: the reconstructed and noise-cleaned detector strain time series, sampled at 16384~Hz.
    \item V1:DQ\_ANALYSIS\_STATE\_VECTOR: the state vector, a time series sampled at 1~Hz and  whose bits provide an information about the quality of the $h(t)$ data and the detector state~\cite{bib:ARframesO4}. Bit~10 is called the CAT1 bit and the data are vetoed as soon as this bit is zero (meaning that data quality is not good for any data analysis).
    \item V1:Hrec\_hoftRepro1AR\_U02\_lastWriteGPS: time series sample at 1~Hz providing the GPS time of the last frame where residual bias and uncertainties channels have been written.
    \item V1:Hrec\_hoftRepro1AR\_U02\_*: a set of frequency vectors stored every 1000~s and providing the frequency-dependent residual bias and uncertainties on the amplitude and phase (in radians) of the Virgo detector strain $h(t)$. The Virgo convention is to provide the values of $h_{rec}/h_{true}$.
    They are:
    \begin{itemize}
      \item two channels to describe the residual bias in amplitude and phase (see section~\ref{sec:hrec:bias}):
      \begin{itemize}
          \item V1:Hrec\_hoftRepro1AR\_U02\_mag\_bias
          \item V1:Hrec\_hoftRepro1AR\_U02\_phase\_bias 
      \end{itemize}
      \item four channels to describe the uncertainties in amplitude and phase (see section~\ref{sec:bias_permmoni}):
      \begin{itemize}
          \item V1:Hrec\_hoftRepro1AR\_U02\_mag\_minus1sigma
          \item V1:Hrec\_hoftRepro1AR\_U02\_mag\_plus1sigma
          \item V1:Hrec\_hoftRepro1AR\_U02\_phase\_minus1sigma
          \item V1:Hrec\_hoftRepro1AR\_U02\_phase\_plus1sigma 
      \end{itemize}
    \end{itemize}
\end{itemize}

More details about the content and conventions used for the AR frames can be found in~\cite{bib:ARframes_prod}.

It is important here to note the convention used to provide the bias and uncertainty in the LVK Collaboration during~O4. 
The information given for the Virgo strain data is the ratio of the reconstructed strain over the expected true value, $h_{rec}/h_{inj}$, as explained in the previous sections. This is the inverse of the information given for the LIGO and KAGRA strain data. This must be taken into account in the data analyses using such information as input, in particular when estimating the astrophysical parameters of the GW sources \cite{bib:GWTC5_methods_orPEpaper}.

\subsection{Production of the AR frame files and validation checks}

While the online $h(t)$ time series was intended to be the final $h(t)$ data product for offline use, we set its values to~0 in the AR frames when the interferometer is not in science observing mode or when the data quality is flagged as bad.
The data quality information was checked by the Virgo detector characterization team and updated in the DQSEGDB database \cite{bib:DQSEGDB} (for instance checking for bad data quality periods not flagged online or, conversely, upgrading to good data quality some periods wrongly flagged online as bad quality). As a consequence, an updated state vector was produced and copied into the AR frames.\\

In addition, it was also necessary to reprocess the $h(t)$ time series whenever it was not available or corrupted, or if not properly reconstructed online. Whenever possible, the cadence for the AR frame production has been chosen to be five weeks, with boundaries corresponding to weekly interferometer maintenance periods. For most of these time periods, the online $h(t)$ time series were valid and thus not reprocessed, except the first weeks of O4b (10 April 2024 to 14 May 2024) for which $h(t)$ was reprocessed.\\

An update of the information about the residual bias and the uncertainties was also required, using the best estimate done from an offline calibration analysis, to replace the preliminary estimation provided in the online frames.
The analysis described in section~\ref{sec:hrecerrors} was run offline on the calibration data of each time period to estimate the residual bias and the uncertainties of $h(t)$. This updated information was then written into the AR frame files.\\

For the O4b run, a first set of AR frame files was ready in June 2025 but errors were found and a new set of AR frame files, for the entire O4b period, was produced and distributed to the LVK Collaboration in October 2025.
For the O4c run, the AR frame files were produced faster and with more systematic validation checks. The AR frame files for the last months of O4c were produced, validated and distributed to the LVK collaboration less than three months after data taking.\\

A set of systematic checks were done on the AR frame files to confirm the presence of all the relevant channels and to check their content. 
For instance, it was verified that $h(t)$ is zero when the state vector indicates bad data quality, and that $h(t)$ is identical to the online $h(t)$ otherwise (except when $h(t)$ was reprocessed). 
We also checked that the updated residual bias and uncertainties were present and with the expected values in the AR frames.
After validation of all the checks, the AR frame files were copied, with checksums, into the final directory for long-term storage.

 \subsection{Summary of the Virgo detector strain residual bias and uncertainties during O4b and O4c}

 During O4b and O4c, the residual bias and the uncertainties on the Virgo detector strain have been estimated offline  over periods of about five weeks. 

\begin{figure}[!ht]
    \centering
	\includegraphics[trim={0 0cm 0 0cm},clip,scale=0.45]{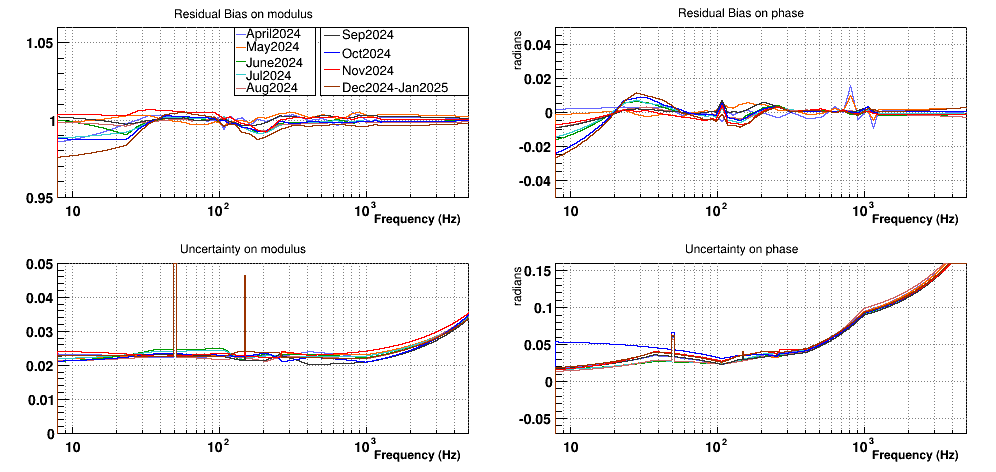} 
    \caption{Residual bias and uncertainties of the Virgo detector strain time series (defined as $h_{rec}/h_{inj}$) for the different periods of O4b, as provided in the AR frame files.
    }
    \label{fig:hrec_bias_diff_o4b}
\end{figure}

\begin{figure}[!ht]
    \centering
	\includegraphics[trim={0 0cm 0 0cm},clip,scale=0.45]{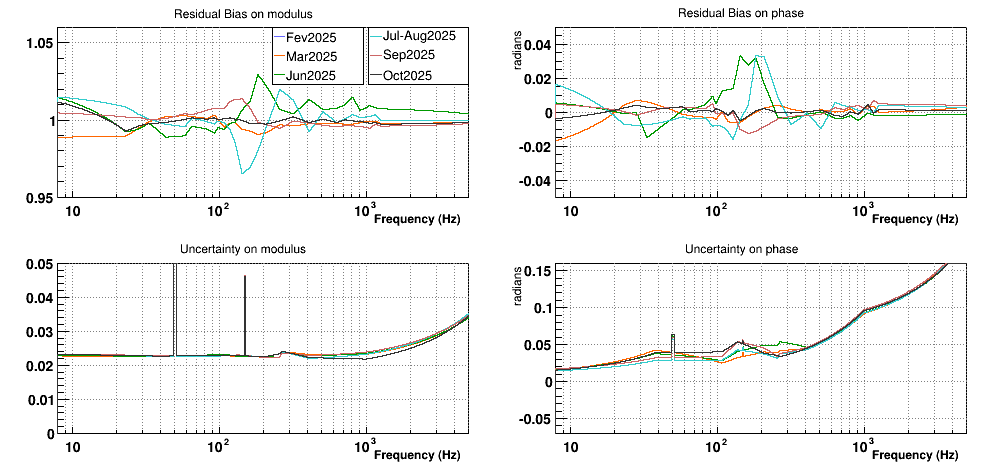} 
    \caption{Residual bias and uncertainties of the Virgo detector strain time series (defined as $h_{rec}/h_{inj}$) for the different periods of O4c, as provided in the AR frame files.
    As stated in the text, the uncertainty on the phase around 50~Hz is underestimated and should be of the order of 500~mrad.
    }
    \label{fig:hrec_bias_diff_o4c}
\end{figure}

Figures~\ref{fig:hrec_bias_diff_o4b} and~\ref{fig:hrec_bias_diff_o4c} summarize the information provided in the AR frame files for the different periods considered (nine periods during O4b and six periods during O4c).
Most of the time, the residual bias remained below 1\% and 20~mrad, except below 30~Hz where larger variations from one month to another are visible. 
The bias was larger during two periods of O4c (June and July 2025), following changes on the interferometer configuration and before the bias subtracted online was updated in the $h(t)$ reconstruction process.
The uncertainties remained very stable during the whole O4b+O4c period. More information about how to use the Virgo uncertainties and residual bias can be found in \cite{bib:lvk_uncertainties} and \cite{bib:GWOSCpaper}.


\section{Conclusion}
The Virgo detector participated in the LVK O4 observing run during the O4b and O4c periods, from 10 April 2024 to 18 November 2025, delivering online $h(t)$ with a latency below 10 seconds and achieving a BNS range of approximately 55~Mpc.\\

For the first time, the Newtonian calibrator, with its systematic uncertainty of about 0.12\%, was used as the reference for the Virgo detector calibration. The calibration of the Photon calibrators (at the 0.48\% level) was adjusted to match the Newtonian calibrators. The NE and WE Photon calibrators were then used to calibrate the mirror actuators,
achieving a precision of 0.77\% in modulus, as summarized in Tables~\ref{tab:MirUncertainties} and \ref{tab:MarUncertainties}.\\

Two sets of Virgo detector strain $h(t)$ time series were produced during O4:
the online version, with a 10~s latency, and the Analysis Ready version, with a latency of a few months. For the first time, the measured, and rather stable, bias of the reconstructed strain was corrected online, resulting in a negligible residual bias in the final Virgo $h(t)$ time series.
For future detections with higher SNR, this procedure will ensure that the low-latency GW searches will not suffer any loss of detection efficiency due to the residual bias.
For O4, this already made the low-latency analyses (searches and spatial localization) simpler, as they can afford ignoring the calibration errors.

Also for the first time, the uncertainty on the Virgo $h(t)$ time series was provided as a frequency-dependent quantity. A frequency-dependent bias estimate was also provided. This information was stored in the same frame files as the $h(t)$ time series, following the frame format \cite{bib:FrameFormat}. This approach allows users to access all relevant information from a single data product. Its main drawback, however, is that even minor update of the uncertainty estimates require the production of a new set of files.

The uncertainty of the $h(t)$ time series, illustrated for a representative period in Fig.~\ref{fig:O4_uncertainty}, has been quite stable throughout O4b and O4c. It was typically
at the level of 2.5\% in modulus, below 40~mrad in phase and approximately 5\,\mus\ in timing.
These values are well within the calibration requirements \cite{bib:LIGO_requirements} established for the Advanced LIGO and Advanced Virgo+ detectors for the O4 run, based on their expected sensitivities. The Virgo frequency-dependent uncertainties on $h(t)$ are comparable to, although slightly larger than, those reported by LIGO~\cite{bib:GWOSC_O4}.
For O4, the estimation of the source parameters was not affected by the Virgo calibration uncertainties.
\\

The Virgo AR frame files have been used by the members of the LVK Collaboration for the scientific analyses and publications based on the O4b and O4c runs. These files are publicly available from the Gravitational Wave Open Science Center (GWOSC) \cite{bib:GWOSC,bib:GWOSCpaper} after each data release (May 2026 for the O4b data set and December 2026 for the O4c data set).\\

Finally, we must remind that the convention used by Virgo to provide the $h(t)$ uncertainty information is the inverse of that adopted by LIGO and KAGRA. A common convention across the detector network is planned for future observing runs, with an implementation expected after 2027.



\section*{Acknowledgements}

The authors gratefully acknowledge the Italian Istituto Nazionale di Fisica Nucleare (INFN),  
the French Centre National de la Recherche Scientifique (CNRS), the Netherlands Organization for Scientific Research (NWO) and the Belgian Fonds de la Recherche Scientifique (FRS-FNRS) 
for the construction and operation of the Virgo detector
and the creation and support of the EGO consortium.
The authors also gratefully acknowledge research support from these agencies as well as by 
the Spanish Agencia Estatal de Investigaci\'on, 
the Consellera d'Innovaci\'o, Universitats, Ci\`encia i Societat Digital de la Generalitat Valenciana and
the CERCA Programme Generalitat de Catalunya, Spain,
the National Science Centre of Poland and the European Union – European Regional Development Fund; Foundation for Polish Science (FNP),
the Polish Minister of Science,
the Hungarian Scientific Research Fund (OTKA),
the French Lyon Institute of Origins (LIO),
the Belgian Fonds de la Recherche Scientifique (FRS-FNRS), 
Actions de Recherche Concertées (ARC) and
Fonds Wetenschappelijk Onderzoek – Vlaanderen (FWO), Belgium,
the Aristotle University of Thessaloniki (AUTH),
the European Commission.
The authors gratefully acknowledge the support of the NSF, STFC, INFN, CNRS and Nikhef for provision of computational resources.

C. Grimaud acknowledges support from the French Agence Nationale de la Recherche for the project ACALCO (ANR-21-CE31-0024).


\section*{References}

\bibliographystyle{iopart-num}
\bibliography{references}

\newpage
\section*{Appendix}

\subsection{Actuators models parameters}
\label{sec:actmodels}
What we refer to as an Advanced Virgo+ mirror actuator
is represented by a transfer function that converts any input voltage into a displacement of the mirror. It is characterized by the following set of parameters:
\begin{itemize}
    \item G : an overall gain expressed in units of meter/Volt
    \item $\delta$ : a frequency-independent delay corresponding to the time interval between the application of the voltage and the resulting mirror motion at very low frequency.
    \item $f_p,Q_p$ : the frequency and quality factor of a double pole associated with the mechanical resonance of the pendulum formed by the mirror and its suspension.
    \item A set of poles and zeros associated with the electronics components of the actuator chain, primarily the coil drivers acting on the magnets attached to the back of the mirror. In practice, this is often reduced to a single pole and a single zero with nearby frequencies denoted $f_{p1}$ and $f_{z1}$.
\end{itemize}

The resulting actuator model can be expressed as the following complex frequency-dependent transfer function:

\begin{equation}
    \large A_{mir}(f) = G \times \exp{(-2\pi jf\delta)} \times \frac{1}{1+jQ_p \frac{f}{f_p}} \times \frac{1 + j\frac{f}{f_{z1}}}{1 + j\frac{f}{f_{p1}}}
\label{eq:actmodel}
\end{equation}

The calibration measurements are performed at several frequencies in 10-1500~Hz range. The resulting set of measured points is then fitted with a model of the form given in Eq.\ref{eq:actmodel}. This fit is done after having normalized the model by the fixed mechanical part (parameters $f_p=0.6$~Hz and $Q_p=1000$).

\end{document}